\documentclass{article}
\usepackage{ijcai20}
\usepackage{times}
\usepackage{soul}
\usepackage{url}
\usepackage[hidelinks]{hyperref}
\usepackage[utf8]{inputenc}
\usepackage[small]{caption}
\usepackage{graphicx}
\usepackage{amsmath}
\usepackage{amssymb}
\usepackage{amsthm}
\usepackage{amsopn}
\usepackage{booktabs}
\usepackage{algorithm}
\usepackage{algorithmic}
\usepackage{enumerate}
\usepackage{color}
\usepackage{multirow}
\usepackage{multicol}
\usepackage{makecell}
\usepackage{hyperref}
\usepackage{diagbox} 
\usepackage{amsthm}
\usepackage{bbding}
\usepackage{amsmath}
\usepackage{graphics}
\usepackage{subfigure}
\usepackage{epsfig}
\usepackage{pifont}
\usepackage{colortbl}
\definecolor{mygray}{rgb}{.955,.965,.99}

\title{Channel-Level Variable Quantization Network for Deep Image Compression}

\author{
	Zhisheng Zhong$^1$ \and Hiroaki Akutsu$^2$ \And Kiyoharu Aizawa$^1$
	\affiliations
	$^1$The University of Tokyo, Japan\\
	$^2$Hitachi Ltd, Japan\\
	\emails{zhisheng@hal.t.u-tokyo.ac.jp,
		hiroaki.akutsu.cs@hitachi.com,
		aizawa@hal.t.u-tokyo.ac.jp}
}

\begin{document}
	
	\maketitle

	\begin{abstract}
		Deep image compression systems mainly contain four components: encoder, quantizer, entropy model, and decoder. To optimize these four components, a joint rate-distortion framework was proposed, and many deep neural network-based methods achieved great success in image compression. However, almost all convolutional neural network-based methods treat channel-wise feature maps equally, reducing the flexibility in handling different types of information. In this paper, we propose a channel-level variable quantization network to dynamically allocate more bitrates for significant channels and withdraw bitrates for negligible channels. Specifically, we propose a variable quantization controller. It consists of two key components: the channel importance module, which can dynamically learn the importance of channels during training, and the splitting-merging module, which can allocate different bitrates for different channels. We also formulate the quantizer into a Gaussian mixture model manner. Quantitative and qualitative experiments verify the effectiveness of the proposed model and demonstrate that our method achieves superior performance and can produce much better visual reconstructions\footnote{ Code address:  \url{https://github.com/zzs1994/CVQN}}.
	\end{abstract}

	\section{Introduction}\label{intro}
	
	Since the development of the internet, increasingly more high-definition digital media data has overwhelmed our daily life. Image compression refers to the task of representing images using as little storage as possible and is an essential topic in computer vision and image processing. 
	
	The typical image compression codecs, e.g., JPEG~\cite{jpeg} and JPEG 2000~\cite{jpeg2000}, generally use some transformations such as discrete cosine transform~(DCT) and discrete wavelet transform~(DWT), which are mathematically well-defined. These compression methods do not fully utilize the nature of data and may introduce visible artifacts such as ringing and blocking. In the last several years, deep learning has revolutionized versatile computer vision tasks~\cite{imagenet,srcnn,resnet}. Image compression based on deep learning, or deep image compression for brevity, has become a popular area of research, which can possibly explore the use of the nature of images beyond conventional compression methods.
	
	A deep image compression system is similar to the conventional one, as it usually includes four components, i.e., encoder, quantizer, entropy model, and decoder, to form the final codec. To train a deep image compression system, a rate-distortion trade-off loss function: ${R} + \lambda {D}$ was proposed in \cite{end2end}, where  $\lambda$ is the balanced hyper-parameter. The loss function includes two competing terms, i.e., ${R}$ measures the bitrate of the final compressed code, and ${D}$ measures the distortion between the original and reconstructed images. Recently, to improve the performance of the deep image compression system further, researchers proposed many novel and effective derivatives for the above four components.

	\paragraph{En/Decoder.} The most popular architecture for en/decoder is based on convolutional neural network~(CNN). E.g., \cite{end2end} proposed a three convolutional layers en/decoder and generalized divisive normalization~(GDN) for activation. \cite{weighted} proposed a nine convolutional layers en/decoder with the residual block~(\cite{resnet}). Google Inc presented three variants (\cite{icrnn,icrnn2,icrnn3}) of a recurrent neural network~(RNN)-based en/decoder to compress progressive images and their residuals. \cite{icgan} proposed a generative adversarial network~(GAN)-based en/decoder for extremely low bitrate image compression, which achieved better user study results.

	\begin{figure*}[t!]
		\centering
		\includegraphics[width=0.92\textwidth]{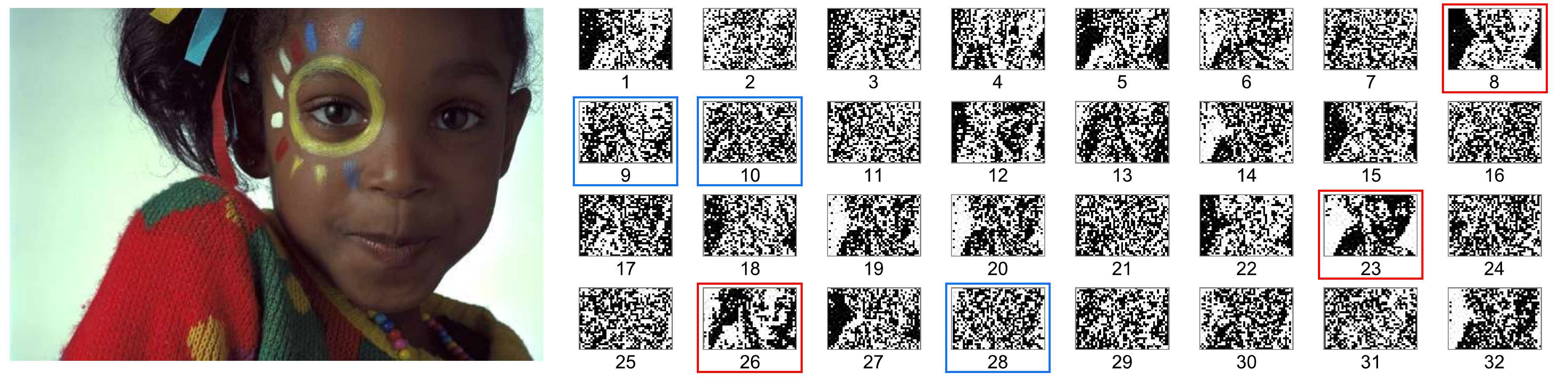} \\
		\vspace{-5pt}
		\begin{multicols}{2}
			\centering 
			\includegraphics[width=0.45\textwidth]{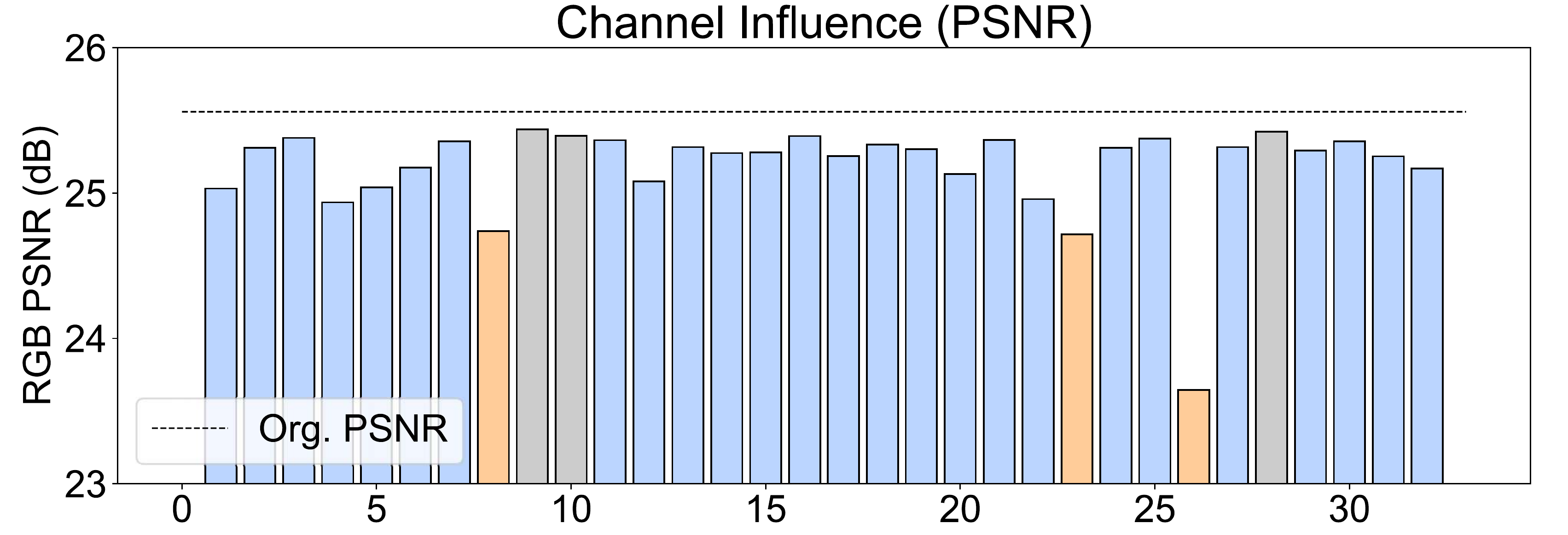} \\ 
			\includegraphics[width=0.45\textwidth]{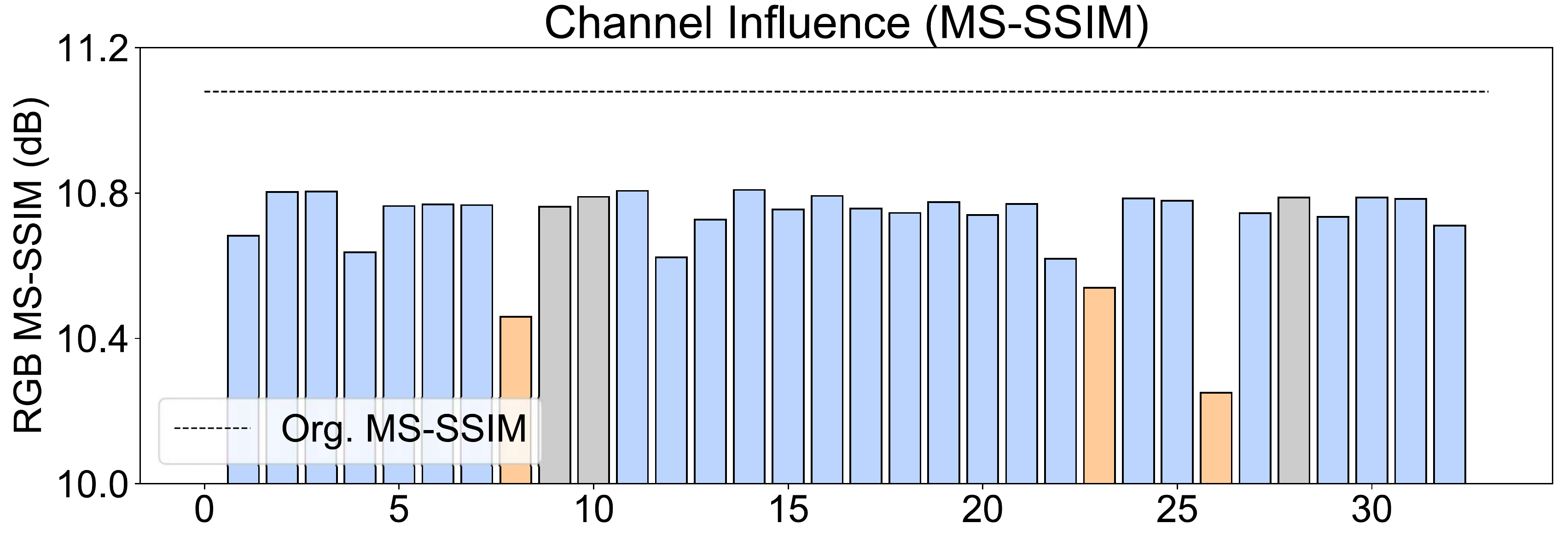} \\
		\end{multicols}
		\vspace{-15pt}
		\caption{Illustration of channel influences. Top left: The original example image (\textit{kodim15}) from the Kodak dataset. Top right: The visual results of the quantized feature map (channel by channel, 32 channels in total). Bottom left: PSNR loss of each channel.  Bottom right: MS-SSIM loss of each channel in decibels: $-10\log_{10}(1-$MS-SSIM$)$. \textit{Best viewed on screen.}}
		\vspace{-15pt}
		\label{fig:channel}
	\end{figure*}
	
	\vspace{-2pt}
	\paragraph{Quantizer.} In conventional codecs, the quantization operation is usually implemented by the round function. However, the gradient of the round function is almost always zero, which is highly unsuitable for deep compression. Thus, many differentiable quantization approaches were proposed by researchers. In \cite{icrnn}, Bernoulli distribution noise was added to implement the map function from continuous values to the fixed set $\{-1,+1\}$. The importance map was proposed in \cite{weighted} to address the spatial inconsistency coding length. Based on the K-means algorithm, the soft quantizer~\cite{conditional} was proposed for the multi-bits quantization case. \cite{end2end} proposed uniformed additive noise for infinite range quantization, whose quantization level is undetermined.
	
	\vspace{-2pt}
	\paragraph{Entropy model.} To further compress the quantized code by through entropy coding methods, e.g., Huffman coding and arithmetic coding, the entropy model was proposed to regularize the redundancy among the quantization code. Almost all entropy arithmetic coding models are motivated by the context-based adaptive binary arithmetic coding~(CABAC) framework. Specifically, in \cite{icrnn2}, they used PixelRNN~\cite{pixelrnn} and long short term memory~(LSTM) architectures for entropy estimation. In \cite{conditional}, they utilized a 3D convolutional model to generate the conditional probability of the quantized code. In \cite{vae_entropy,minnen2018joint,lee2018context}, they proposed a variational autoencoder~(VAE) with a scale hyperprior to learn the context distribution, which achieves consequently achieving better compression results.
	
	\vspace{-5pt}
	\section{Channel Influences}\label{channel_influence}
	\vspace{2pt}
	
	All previous deep image compression systems view all channels as a unified whole and ignore the channel-level influences. However, useful information is unevenly distributed across channels. Channel redundancy and uneven distribution have been widely studied in the field of network compression~\cite{thinet,slim,he2017channel}. In this study, we utilize a toy example model to illustrate its feasibility in deep image compression. We use a simple encoder and quantizer to extract features and quantize them. The final quantized feature map has $32$ channels. We allocate one bit for quantization, i.e., its quantization level is two. Evaluating on the Kodak dataset, this toy model yields an average MS-SSIM~\cite{msssim} of 0.922 at an average rate of 0.112 bits per pixel~(BPP). In the top part of Fig.~\ref{fig:channel}, we present the visual results of the quantized feature map (channel by channel, 32 channels) by using \textit{kodim15} from Kodak. The visual results indicate that Channel-8, 23, and 26 have similar content and profile (similar to low-frequency information) with the original image. By contrast, some visualizations,  e.g., Channel-9, 10, and 28 appear disorganized and could not be recognized (similar to high-frequency information). We also make quantitative comparisons. We conduct 32 experiments. In each experiment, we cut one relative channel (set its values to 0) of the quantized feature map to observe the influence of each channel on the final reconstruction results. The bottom of Fig.~\ref{fig:channel} depicts the PSNR loss of each channel on the left and the MS-SSIM loss of each channel on the right. Consistent with the analysis of visual results, Channel-8, 23, and 26 are significant for reconstruction, whereas Channel-9, 10, and 28 are negligible. Moreover, this phenomenon appears on all images of the dataset. Thus, the problem is as follows: Can we design a variable deep image compression system to ensure the allocation of more bits for important channels and the reduction of bitrate for negligible channels? In this paper, we propose a novel network to solve this issue.
	
	\vspace{2pt}
	The overall contributions of this study are three-fold:
	\vspace{-2pt}
	\begin{itemize}
		\item We analyze the channel influences in deep image compression. We propose a novel variable channel controller to effectively utilize channel diversity. To the best of our knowledge, we are the first to perform image compression in a channel-level manner.
		\vspace{-3pt}
		\item We propose a novel quantizer based on Gaussian mixture model~(GMM). This novel  quantizer has powerful representation and is a more generalized pattern for the existing finite quantizers.
		\vspace{-3pt}
		\item Extensive quantitative and qualitative experiments show that our method achieves superior performance over the state-of-the-art methods without a hyperprior VAE.
	\end{itemize}
	
	\vspace{-10pt}
	\section{Approach}
	\vspace{2pt}
	
	\begin{figure*}[t]
		\begin{center}
			\includegraphics[width=1\linewidth]{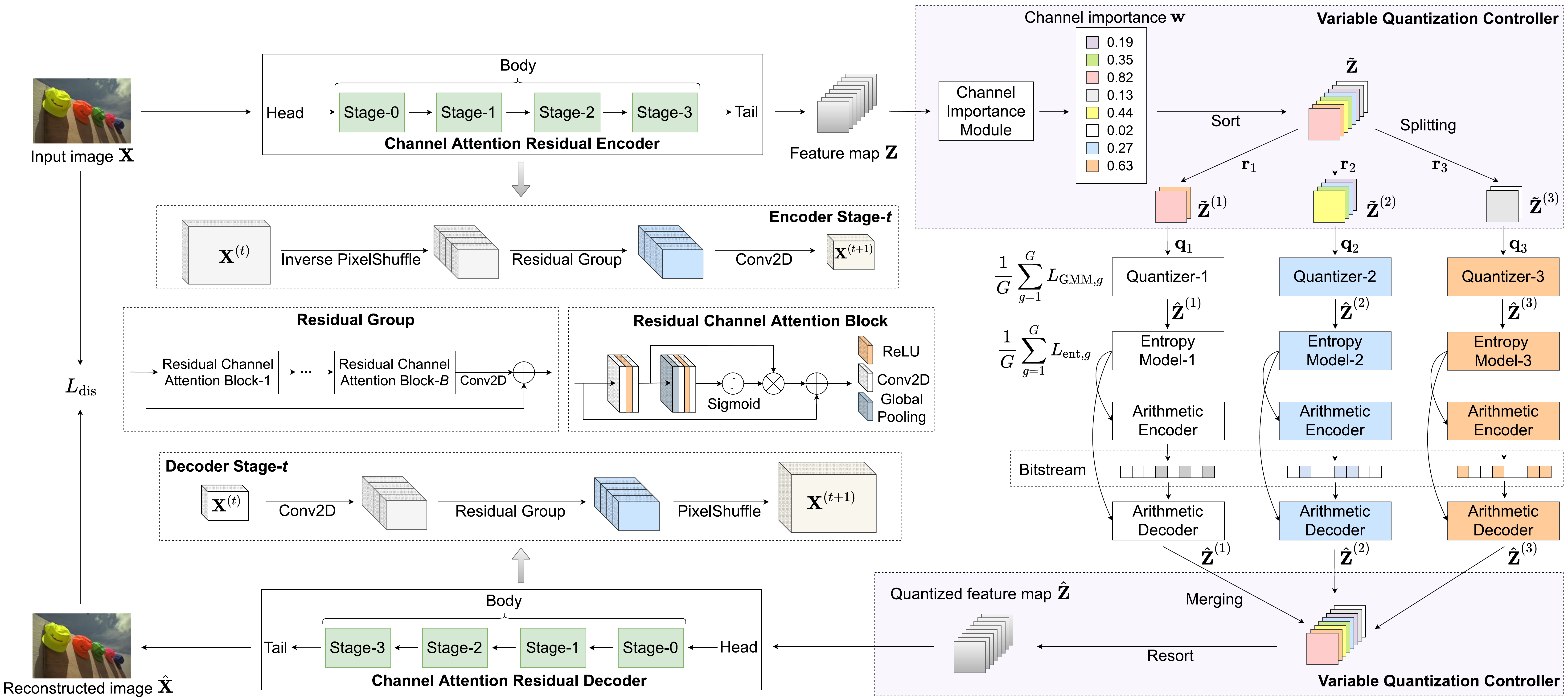}
		\end{center}
		\vspace{-8pt}
		\caption{Framework of the channel-level variable quantization network. The entire encoder and decoder bodies both contain four stages.  $C=8$, $G=3$, and $\mathbf{r}=[25\%, 50\%, 25\%]^\top$ are selected for the illustration of variable quantization controller. \textit{Best viewed on screen.}}
		\label{fig:frame}
		\vspace{-12pt}
	\end{figure*}
	
	\vspace{-2pt}
	The framework of the proposed system is shown in Fig.~\ref{fig:frame}. In this section, we first introduce the channel attention residual network for encoding and decoding. Then, we present a novel quantizer based on GMM. Finally, we illustrate the details of the variable quantization level controller, which makes the entire system able to dynamically alter the quantization levels for each channel.
	
	\vspace{-2pt}
	\subsection{Channel Attention Residual En/Decoder}\label{encoder}

	Our channel attention residual encoder comprises three parts: head, body, and tail. The head module contains one convolutional layer, which transforms the original image into feature map $\mathbf{X}^{(0)}$ with $C_0$ channels. The body of the encoder is shown in left part of Fig.~\ref{fig:frame}. The entire body includes four stages. In each stage, the output feature map is only half the resolution $(h, w)$ of the input feature map. We denote the input feature map at Stage-$t$ as $\mathbf{X}^{(t)}\in \mathbb{R}^{C_t \times H \times W}$. Motivated by the super-resolution task's method~\cite{pixelshuffle}, we use inverse PixelShuffle to implement the down-sampling operation. It can be expressed as:
	\begin{equation}\label{ips}
	\mathcal{IPS}(\mathbf{X}^{(t)})_{c(di+j), h, w} = \mathbf{X}^{(t)}_{c, dh + i, dw + j}, \ 1\leq i,j \leq d,
	\end{equation}
	where $d$ is the down-sampling factor. It is a periodic shuffling operator that rearranges the elements of a $C_t \times H \times W$ tensor to a tensor of shape $d^2C_t \times \frac{1}{d}H \times \frac{1}{d}W$. Notably, this operator preserves all information of the input because the number of elements does not vary.  We also found that inverse PixelShuffle can improve the stability of training and reduce the memory costs relative to the down-sampling convolution.

	Previous CNN-based image compression methods treat channel-wise features equally, which is not flexible for real cases. To make the network focus on more informative features, we follow \cite{senet,rcan,srcliquenet} and exploit the inter-dependencies among feature channels. We send the feature map to the residual group module, shown in left part of Fig.~\ref{fig:frame}. The residual group consists of $B$ residual channel attention blocks, which are used to extract the inter-dependencies among feature channels and distill the feature map. The residual group does not change the number of channels.
	
	Finally, we add a convolutional layer to alter the number of channels from $C_t$ to $C_{t+1}$ for the next stage. Thus the output of Stage-$t$ is $\mathbf{X}^{(t+1)}\in \mathbb{R}^{C_{t+1} \times \frac{1}{d}H \times \frac{1}{d}W}$, which is also the input of the next stage.
	
	After four stages of processing in the body, a convolutional layer, appended as the tail part, generates the compressed (latent) representation $\mathbf{Z}$ with $C$ channels, where $C$ can be varied manually for different BPPs. Similarly, the architecture of the decoder is simply the inverse version of the encoder. As shown in left part of Fig.~\ref{fig:frame}, we replace inverse PixelShuffle with PixelShuffle for the up-sampling operation.

	\subsection{GMM Quantizer}\label{gmm}
	For the quantizer, we propose a novel quantization method based on GMM. Concretely, we model the prior distribution $p(\mathbf{Z})$ as a mixture of Gaussian distributions:
	\vspace{-2pt}
	\begin{equation}
	p(\mathbf{Z}) = \prod_{i}\sum_{q=1}^{Q}\pi_q\mathcal{N}(z_i|\mu_q, \sigma_q^2),
	\end{equation}
	where $\pi_q$, $\mu_q$, and $\sigma_q$ are the learnable mixture parameters and $Q$ is the quantization level.
	
	We obtain the forward quantization result by setting it to the mean that takes the largest responsibility:
	\vspace{-1pt}
	\begin{equation}
	\hat{z}_i \leftarrow \mathop{\arg\max}_{\mu_j}\frac{\pi_j\mathcal{N}(z_i|\mu_j, \sigma_j^2)}{\sum_q^Q\pi_q\mathcal{N}(z_i|\mu_q, \sigma_q^2)}.
	\label{eqn:forward}
	\end{equation}
	\vspace{-1pt}
	
	Obviously, Eqn. (\ref{eqn:forward}) is non-differentiable. We relax $\hat{z}_i$ to $\tilde{z}_i$ to compute its gradients during the backward pass by:
	\vspace{-1pt}
	\begin{equation}
	\tilde{z}_i = \sum_{j=1}^Q\frac{\pi_j\mathcal{N}(z_i|\mu_j, \sigma_j^2)}{\sum_q^Q\pi_q\mathcal{N}(z_i|\mu_q, \sigma_q^2)}\mu_j.
	\end{equation}
	\vspace{-1pt}
	
	Unlike the conventional GMM, which optimizes $\pi_q$, $\mu_q$, and $\sigma_q$ by using the expectation maximization~(EM) algorithm, we learn the mixture parameters by minimizing the negative likelihood loss function through the network back-propagation. We denote the prior distribution loss function of GMM quantizer as:
	\vspace{-5pt}
	\begin{equation}
	L_{\rm{GMM}} \!=\! - {\rm{log}}\left(p(\mathbf{Z})\right)  \!= \! - \sum_{i}{\rm{log}}\sum_{q=1}^{Q}\pi_q\mathcal{N}(z_i|\mu_q, \sigma_q^2).
	\end{equation}
	\vspace{-5pt}

	Here, we would like to make a comparison between the GMM quantizer and the soft quantizer~\cite{soft}. The soft quantizer can be viewed as a differentiable version of the K-means algorithm. If the mixture parameters satisfy: $\pi_1=\pi_2=\dots=\pi_Q=1/Q$ and $\sigma_1=\sigma_2=\dots=\sigma_Q=\sqrt{2}/2$, the GMM quantizer will degenerate to the soft quantizer, which implies that the GMM quantizer has a more powerful representation and is a more generalized model.

	\vspace{-1pt}
	\subsection{Variable Quantization Controller}\label{controller}
	
	As mentioned in Sec.~\ref{channel_influence}, each channel of the quantized feature map may have a different impact on the final reconstruction results. To allocate appropriate bitrates for different channels, we propose the variable quantization controller model.
	
	The illustration of the variable quantization controller is shown in the right part of Fig.~\ref{fig:frame}. In the variable quantization controller, there are two key components: channel importance module and splitting-merging module. In the following, we will introduce the mechanism of these two modules in detail.

	\vspace{-1pt}
	\subsubsection{Channel Importance Module}\label{cim}
	\vspace{2pt}
	
	The input of the channel importance module is $\mathbf{Z}$, which is the output of the encoder mentioned in Sec.~\ref{encoder}. Let us denote the channel number of $\mathbf{Z}$ as $C$ ($C = 8$ in Fig.~\ref{fig:frame}). We expect the channel importance module to generate a channel importance vector $\mathbf{w} \in \mathbb{R}^C_+$. Each element $\mathbf{w}_c$ represents the reconstruction importance of Channel-$c$.
	
	Here, we design three types of channel importance module:
	
	\paragraph{Sequeze and excitation block-based.} We utilize average pooling and two convolutional layers to operate $\mathbf{Z}$ (refer~\cite{senet}) and get an $M \times C$ matrix, where $M$ is the mini-batch size. We generate a learnable channel importance vector $\mathbf{w}$ by using the mean operation on the matrix by reducing the first dimension ($M$).
	
	\paragraph{Reconstruction error-based.} We perform three steps to implement it: First, we construct a validation dataset by randomly selecting $N$ images from the training dataset. Second, we prune the $c$-th channel of the $n$-th image's feature map $\mathbf{Z}_{n,c}$: $\mathbf{Z}_n(c,:,:)=0$. Last, we represent $\mathbf{w}_c $ by calculating the average MS-SSIM reconstruction error of each channel over the validation dataset:
	\vspace{-2pt}
	\begin{equation}
	\mathbf{w}_c =  \frac{1}{N}\sum_{n=1}^{N}d_{\rm MS\textendash SSIM} \left(\mathbf{I}_n,{\rm{Dec}} \left({\rm{Qua}}\left(\mathbf{Z}_{n,c}\right)\right)\right),
	\end{equation}
	\vspace{-2pt}
	
	\noindent where $\mathbf{I}_n$ is the $n$-th image of the validation dataset, ${\rm{Dec}}$ and ${\rm{Qua}}$ are represent the decoder and quantizer, respectively. 
	
	\vspace{2pt}
	\paragraph{Predefined.} We directly predefine the channel importance vector $\mathbf{w}$ as $\mathbf{w}_c = c$, which is fixed during the training and evaluation process.

	\subsubsection{Splitting-Merging Module}
	\vspace{2pt}
	
	At the beginning of the splitting-merging module, we sort the feature map $\mathbf{Z}$ in ascending order according to the channel importance vector $\mathbf{w}$. Because the new feature map is well organized, we split it to $G$ groups ($G = 3$ in Fig.~\ref{fig:frame}). The $G$ portions of the feature map are quantized and encoded using different quantization levels in different groups.
	
	After the splitting operation, the $C$ channels are divided into $G$ groups. We denote the ratio vector of $G$ groups as $\mathbf{r}$, which satisfies: $\sum_{g}^G\mathbf{r}_g=1$, and $\forall g, \mathbf{r}_g>0$. Here, we use the right part of Fig.~\ref{fig:frame} to explain its mechanism. Suppose that the parameters $C=8$, $G=3$, and $\mathbf{r}=[25\%, 50\%, 25\%]^\top$, Channel-1 and 2 will be assigned Group-1 for quantization and encoding, Channel-3, 4, 5, and 6 will be assigned Group-2 and Channel-7 and 8 will be assigned Group-3. On the other hand, because the channel importances of Channel-1 and 2 are smaller than the others, we use smaller quantization level $\mathbf{q}_1$ for quantizing and encoding. Similarly, we apply a larger quantization level $\mathbf{q}_3$ to quantize and encode Channel-7 and 8. At the last step, we merge $G$ groups and reorder the channel dimension to construct the final compressed result. 
	
	\begin{figure*}[t!]
		\centering
		\begin{multicols}{2}
			\centering \includegraphics[width=0.54\textwidth]{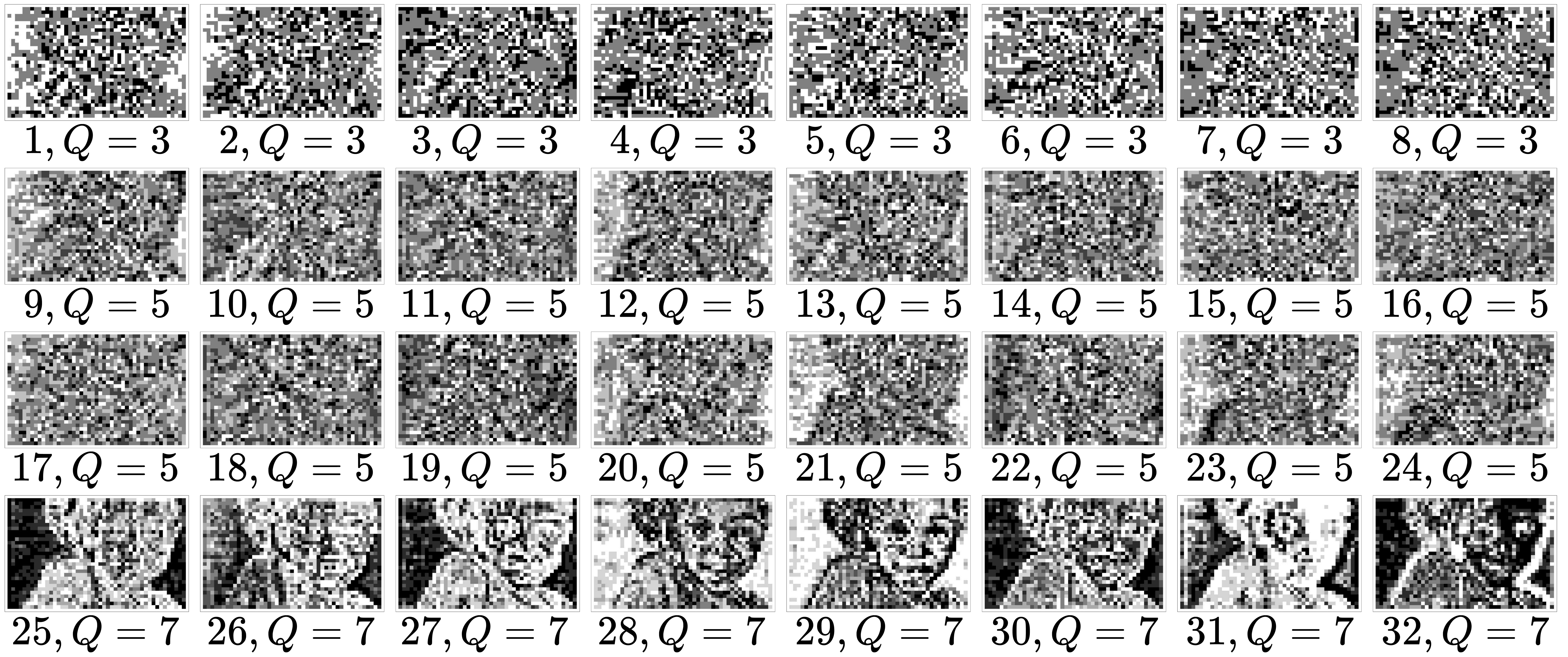} \\\hspace{10pt}
			\includegraphics[width=0.44\textwidth]{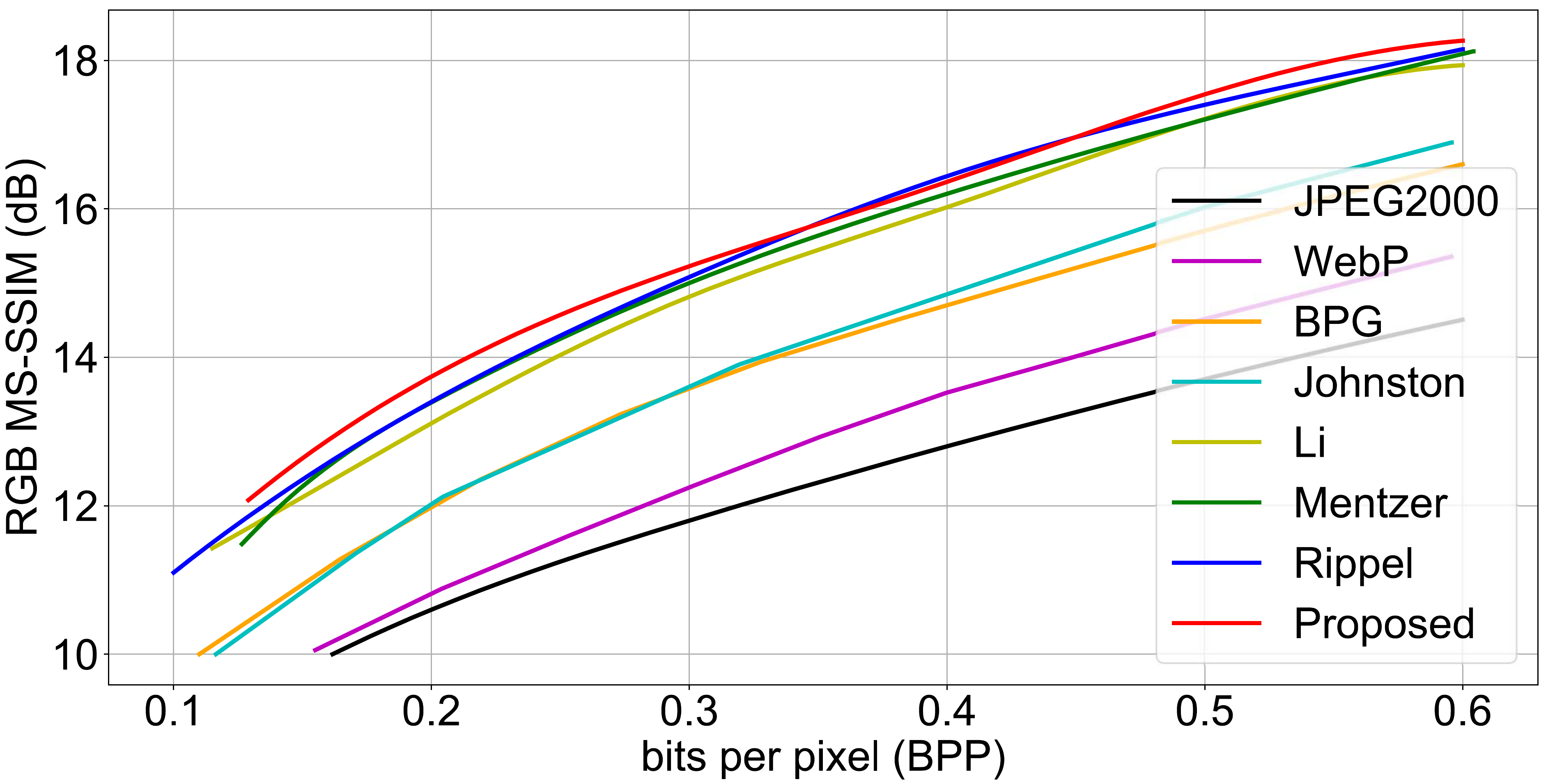}\\ 
		\end{multicols}
		\vspace{-18pt}
		\caption{Left: Visualization results (\textit{kodim15}) of the predefined model's quantized feature map, which contains three quantization levels: 3, 5, and 7. Right: Comparisons of the rate-distortion curves on Kodak. MS-SSIM values are converted into decibels. \textit{Best viewed on screen.}}
		\label{fig:visual_compare}
		\vspace{-15pt}
	\end{figure*}
	
	\vspace{-1pt}
	\subsubsection{Analysis}
	\vspace{2pt}
	
	Here, we conduct an analysis, examining under what condition the variable quantization controller can theoretically guarantee a better compression rate than that of the original one-group model. We suppose that the feature map $\mathbf{Z}$ is a $C \times H \times W$ tensor and the channel number of Group-$g$ is $C_g=C\mathbf{r}_g$. Obviously, it satisfies $\sum_{g}C_g=C$. $\hat{\mathbf{Z}}$ has the same dimensions as $\mathbf{Z}$ because $\hat{\mathbf{Z}}$ is simply the quantized version of $\mathbf{Z}$. Because the number of dimensions ${\dim(\hat{\mathbf{Z}})}$ and the quantization level $Q$ are finite, the entropy is bounded by  $H(\hat{\mathbf{Z}}) \leq \dim(\hat{\mathbf{Z}}){\rm{log_2}}(Q)= CHW{\rm{log_2}}(Q)$ (refer, e.g., \cite{cover2012elements}). Contrastingly, for $G$ groups, suppose that the quantization level vector is $\mathbf{q}=[\mathbf{q}_1, \mathbf{q}_2, ..., \mathbf{q}_G]^\top$, then, the entropy upper-bound of $\{\hat{\mathbf{Z}}_g\}$ is:
	
	\vspace{-5pt}
	\begin{equation}
	H(\{\hat{\mathbf{Z}}_g\})=\sum_{g=1}^{G}H(\hat{\mathbf{Z}}_g)\leq HWC\sum_{g=1}^{G}\mathbf{r}_g{\rm{log_2}}(\mathbf{q}_g).
	\end{equation}
	\vspace{-5pt}
	
	Thus, if the $G$ groups satisfy $\mathbf{r}^\top{\rm{log_2}}(\mathbf{q}) < {\rm{log_2}}(Q)$, the variable quantization controller will provide a lower entropy upper-bound than the conventional one-group model. On the other hand, although $\hat{\mathbf{Z}}$ has the same total number of elements as $\{\hat{\mathbf{Z}}_g\}$, $\hat{\mathbf{Z}}$ has only $Q$ values to pick up, whereas $\{\hat{\mathbf{Z}}_g\}$ has $\sum_{g}\mathbf{q}_g$ values, indicating that $\{\hat{\mathbf{Z}}_g\}$ may have better diversity.

	Overall, in the variable quantization controller, we choose the GMM quantizer (in Sec.~\ref{gmm}) and the 3D CNN-based context model (refer \cite{conditional}) for quantization, and entropy estimating, respectively. All quantized feature maps $\{\hat{\mathbf{Z}}_k\}$ will concatenate together and be sent to the decoder. The final loss function of the entire system becomes:
	
	\vspace{-10pt}
	\begin{equation}\label{eqn:loss}
	L =\alpha L_{\rm{dis}}  + \frac{1}{G}\sum_{g=1}^{G}L_{{\rm{ent}},g} + \beta \frac{1}{G}\sum_{g=1}^{G}L_{{\rm{GMM}},g}.
	\end{equation}
	
	\vspace{5pt}
	\section{Experiments}
	\vspace{2pt}
	\subsection{Implementation and Training Details}
	\vspace{2pt}
	\paragraph{Datasets.} We merge three common datasets, namely DIK2K~\cite{div2k}, Flickr2K~\cite{flickr}, and CLIC2018, to form our training dataset, which contains approximately 4,000 images in total. Following many deep image compression methods, we evaluate our models on the Kodak dataset with the metrics MS-SSIM for lossy image compression.
	
	\renewcommand\arraystretch{1.1}
	\begin{table}[t]
		\centering
		\vspace{5pt}
		\begin{tabular}{p{1.4cm}<{\centering}p{1.4cm}<{\centering}p{1.45cm}<{\centering}p{2.55cm}<{\centering}}
			\hline 
			$\mathbf{q}$ & CI Type   & MS-SSIM & BPP \\
			\hline \hline
			$[5]$ & None   & 0.9651 & 0.2664 \\
			\rowcolor{mygray}
			$[3, 5, 7]^\top$ & SE-based & 0.9646 & 0.2608 ($\downarrow$ 2.11\%) \\
			$[3, 5, 7]^\top$& RE-based & 0.9652 &  0.2586 ($\downarrow$ 2.93\%) \\
			\rowcolor{mygray}
			$[3, 5, 7]^\top$ & Predefine & \textbf{0.9653} & \textbf{0.2576 ($\downarrow$ 3.31\%)} \\
			\hline 
		\end{tabular}
		\vspace{-5pt}
		\caption{Investigation of channel importance module. We run it three times and show the average results. CI Type denotes the type of channel importance module mentioned in Sec.\ref{cim}. }
		\vspace{-10pt}
		\label{table:variable}
	\end{table}

	\vspace{-1pt}
	\paragraph{Parameter setting.} In our experiments, we use the Adam optimizer~\cite{adam} with a mini-batch size $M$ of 32 to train our five models on $256\times256$ image patches. We vary the quantized feature map $\hat{\mathbf{Z}}$'s channel number $C$ from 16 to 80 to obtain different BPPs. The total number of training epochs equals to $400$. The initialized learning rates are set to $1 \times 10^{-4}, 1 \times 10^{-4}, 5 \times 10^{-5}$ and $1 \times 10^{-4}$ for the encoder, quantizer, entropy model, and decoder, respectively.  We reduce them twice (at Epoch-200 and Epoch-300) by a factor of five during training. In the channel attention residual en/decoder, we set the number of residual channel attention blocks $B=6$ for all stages. The channel numbers for each stage in the encoder are 32, 64, 128, and 192, respectively, whereas those for each stage in the decoder are 192, 128, 64, and 32, respectively. In the variable quantization controller, we set the number of groups $G=3$. The ratio vector $\mathbf{r}=[25\%, 50\%, 25\%]^\top$. For loss function Eqn.~(\ref{eqn:loss}), we choose negative MS-SSIM for the distortion loss $L_{\rm{dis}}$ and $\alpha=128$; we select cross entropy for the entropy estimation loss $L_{\rm{ent}}$ and $\beta=0.001$.

	\vspace{-2pt}
	\subsection{Ablation Study}
	\vspace{1pt}

	\begin{table}[t]
		\renewcommand\arraystretch{1.1}
		\centering
		
		\setlength{\tabcolsep}{0.3mm}{
			\vspace{5pt}
			\begin{tabular}{p{1.8cm}<{\centering}p{1.8cm}<{\centering}p{1.8cm}<{\centering}p{2.5cm}<{\centering}}
				\hline 
				$\mathbf{q}$      & PSNR   & MS-SSIM & BPP \\
				\hline \hline		
				$[5]$   & 27.926 & 0.9651 & 0.2664 \\
				\rowcolor{mygray}
				$[4, 5, 6]^\top$ & 28.012 & 0.9652  & 0.2639($\downarrow$ 0.94\%) \\
				$[3, 5, 7]^\top$  & \textbf{28.024} &\textbf{0.9653}& 0.2576($\downarrow$ 3.31\%) \\
				\rowcolor{mygray}
				$[2, 5, 8]^\top$ & 27.982& 0.9644  & \textbf{0.2471($\downarrow$ 7.24\%)} \\
				\hline 
		\end{tabular}}		
		\vspace{-4pt}
		\caption{Investigation of the combination in $\mathbf{q}$. We run it three times and show the average results.}
		\vspace{-13pt}
		\label{table:combination}
	\end{table}

	\begin{figure*}
		\centering
		\subfigure{
			\begin{minipage}[b]{0.185\linewidth}
				\centering {\small \bf Org.}\vspace{1pt}
				\includegraphics[width=1\linewidth]{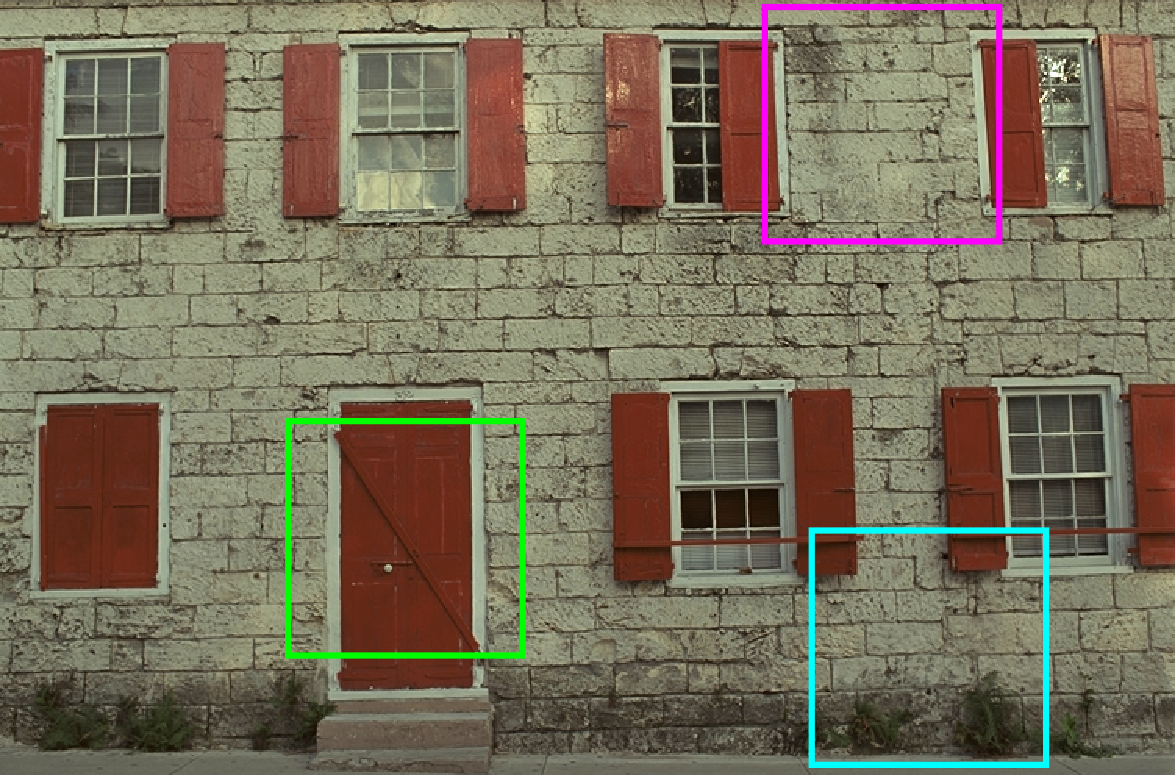}\vspace{2pt}
				\includegraphics[width=1\linewidth]{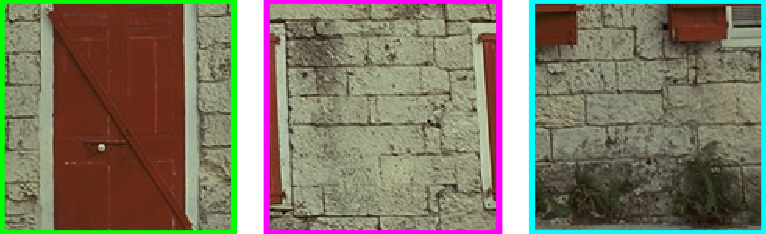}\vspace{-1pt}
				\centering {\small MS-SSIM / BPP}\vspace{4pt}
				\includegraphics[width=1\linewidth]{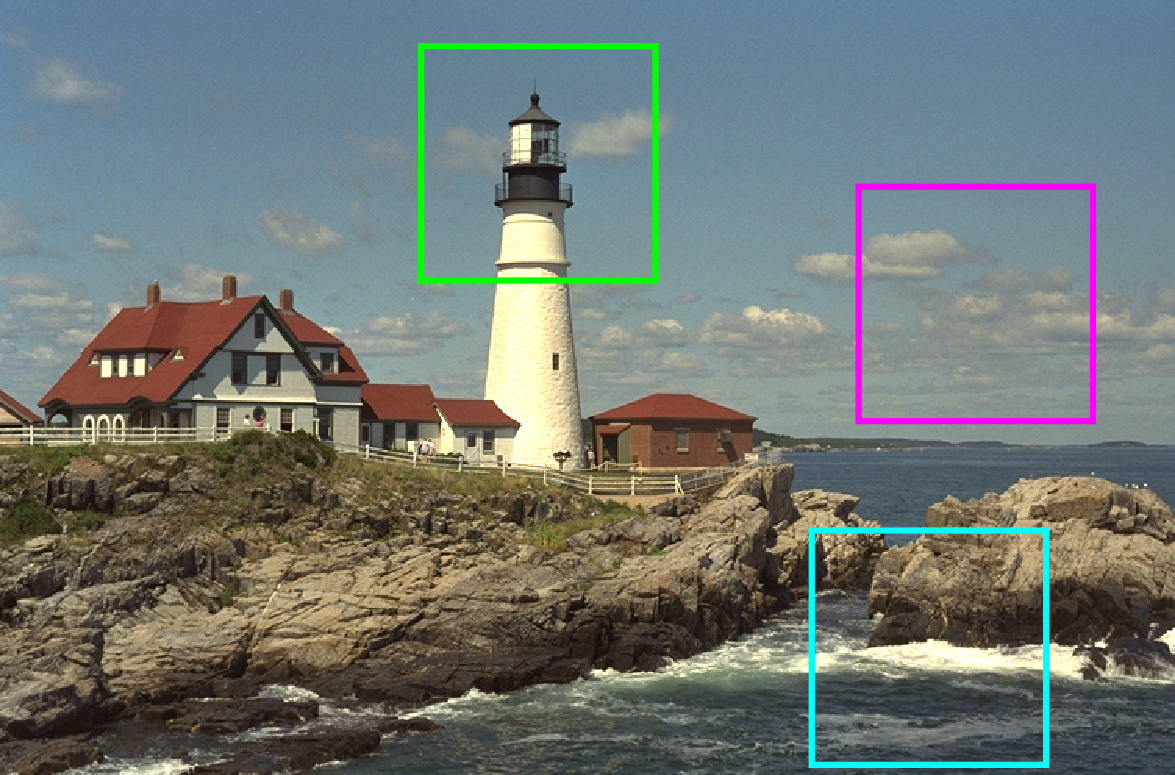}\vspace{2pt}
				\includegraphics[width=1\linewidth]{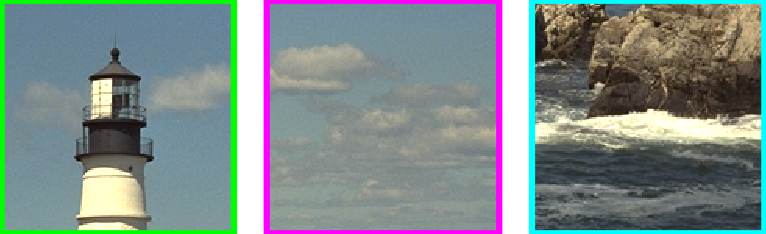}\vspace{-1pt}
				\centering {\small MS-SSIM / BPP}
		\end{minipage}}
		\subfigure{
			\begin{minipage}[b]{0.185\linewidth}
				\centering {\small \bf WebP}\vspace{1pt}
				\includegraphics[width=1\linewidth]{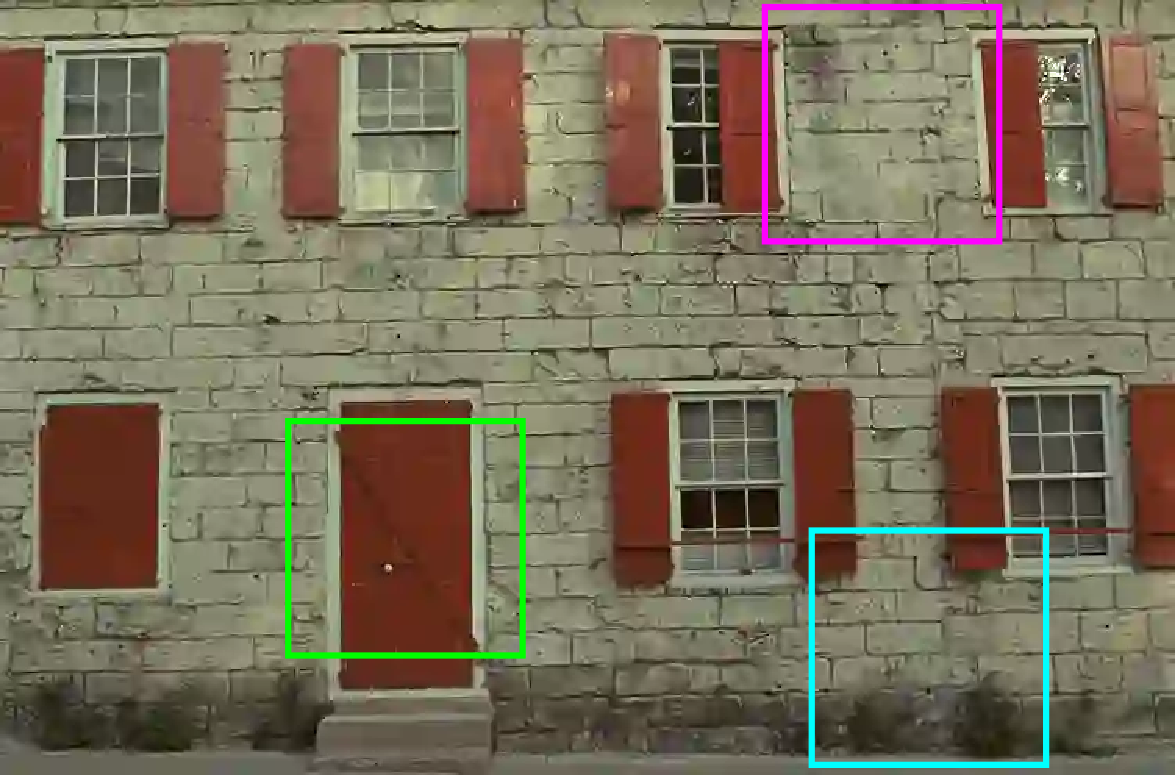}\vspace{2pt}
				\includegraphics[width=1\linewidth]{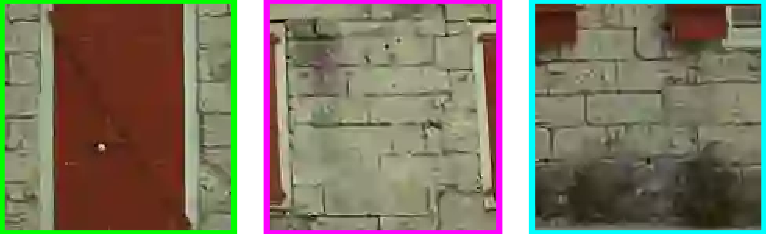}\vspace{-1pt}
				\centering {\small 0.903 / 0.250}\vspace{4pt}
				\includegraphics[width=1\linewidth]{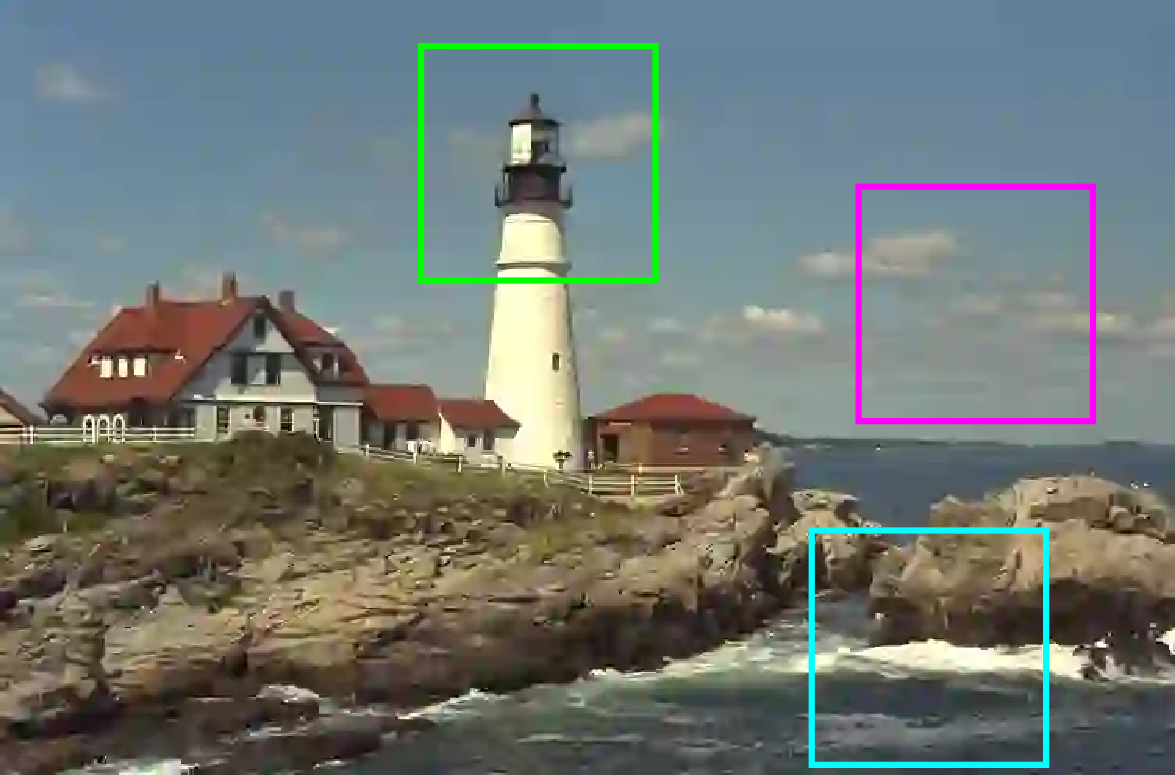}\vspace{2pt}
				\includegraphics[width=1\linewidth]{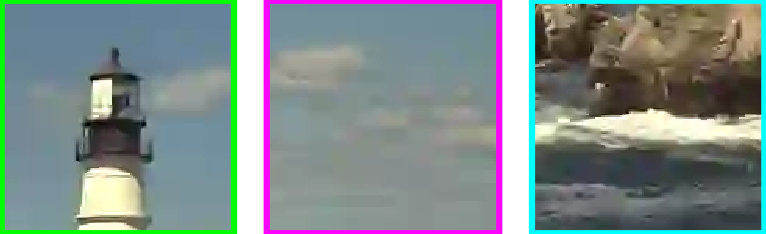}\vspace{-1pt}
				\centering {\small 0.918 / 0.160}
		\end{minipage}}
		\subfigure{
			\begin{minipage}[b]{0.185\linewidth}
				\centering {\small \bf BPG}\vspace{1pt}
				\includegraphics[width=1\linewidth]{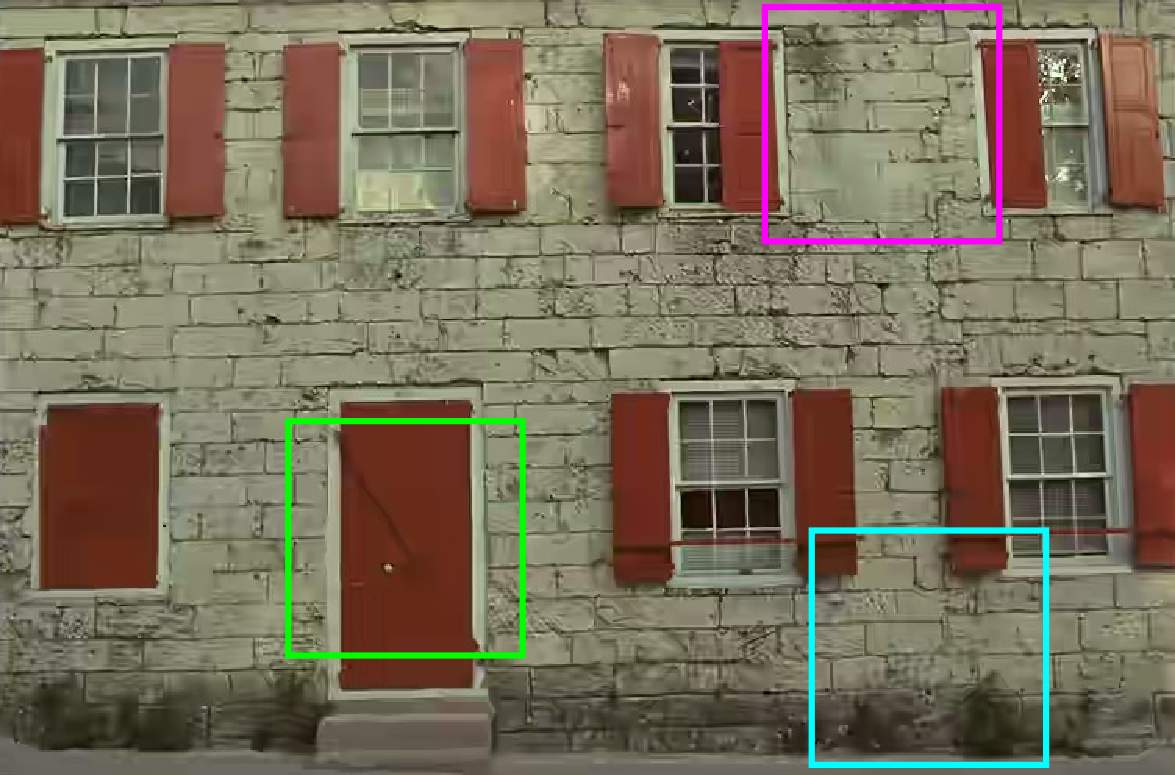}\vspace{2pt}
				\includegraphics[width=1\linewidth]{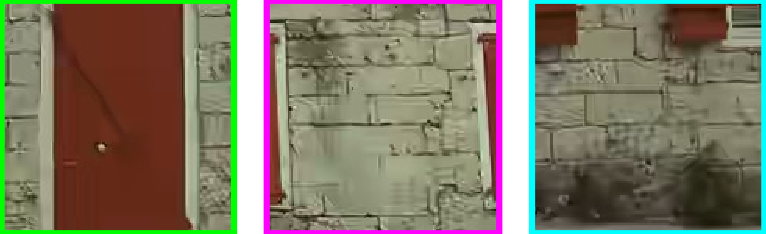}\vspace{-1pt}
				\centering {\small 0.927 / 0.246}\vspace{4pt}
				\includegraphics[width=1\linewidth]{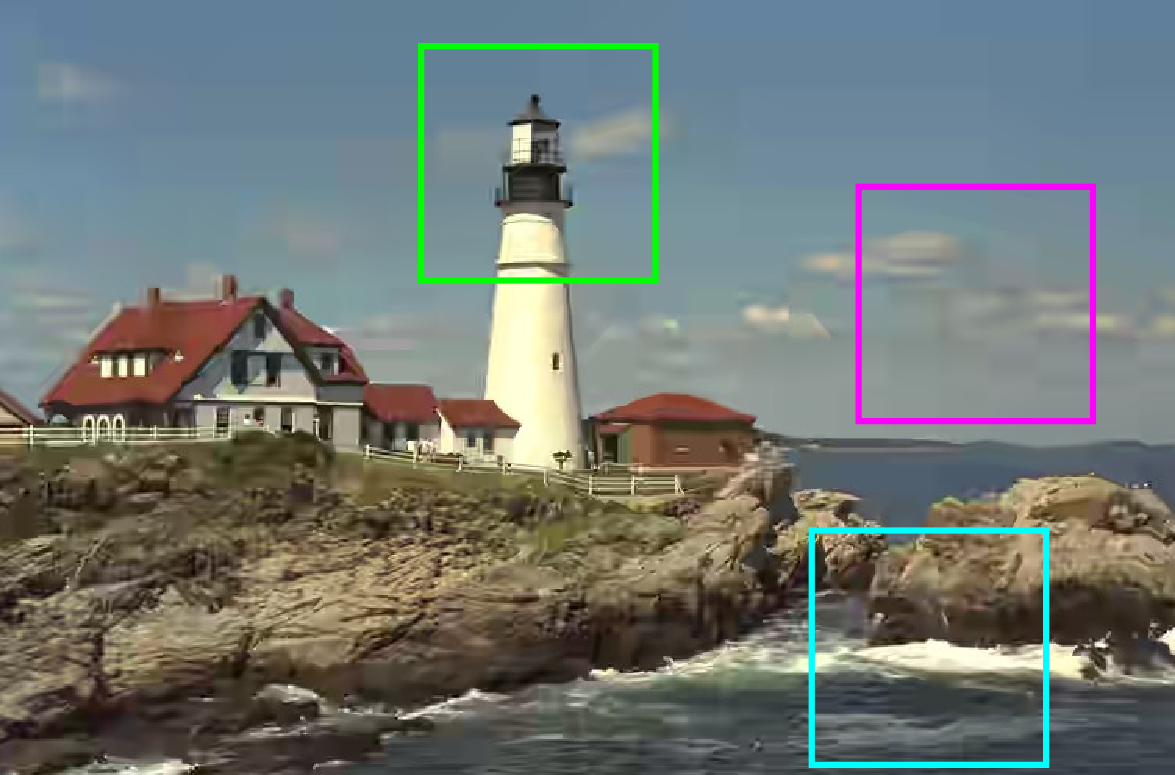}\vspace{2pt}
				\includegraphics[width=1\linewidth]{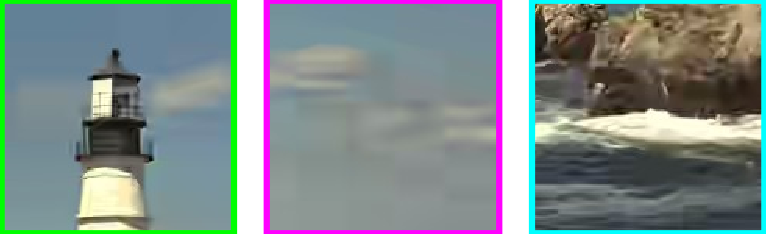}\vspace{-1pt}
				\centering {\small 0.931 / 0.137}
		\end{minipage}}
		\subfigure{
			\begin{minipage}[b]{0.185\linewidth}
				\centering {\small \bf Mentzer et al.}\vspace{1pt}
				\includegraphics[width=1\linewidth]{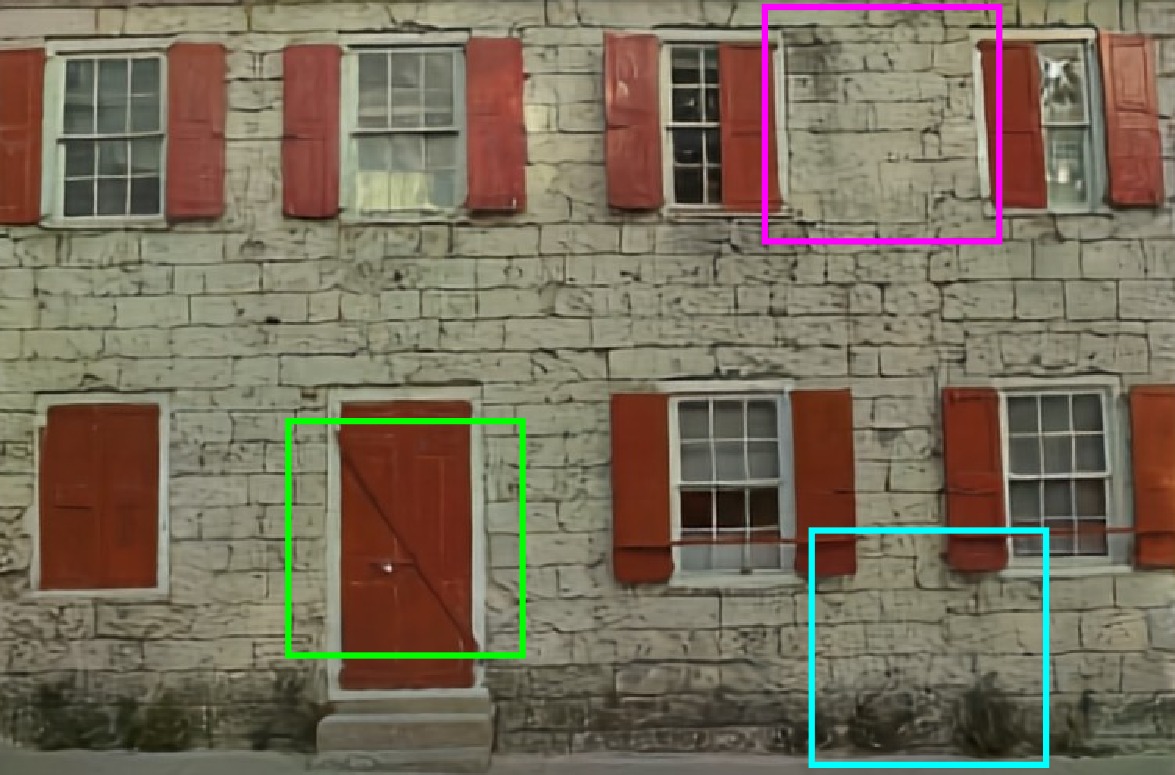}\vspace{2pt}
				\includegraphics[width=1\linewidth]{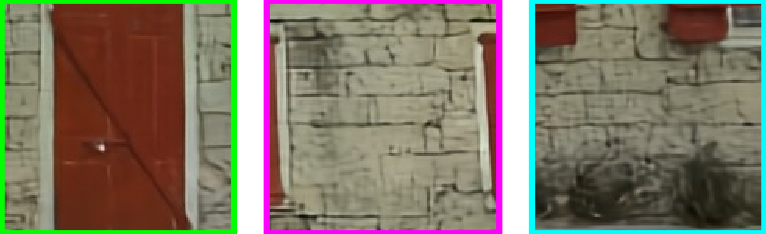}\vspace{-1pt}
				\centering {\small 0.940 / 0.239}\vspace{4pt}
				\includegraphics[width=1\linewidth]{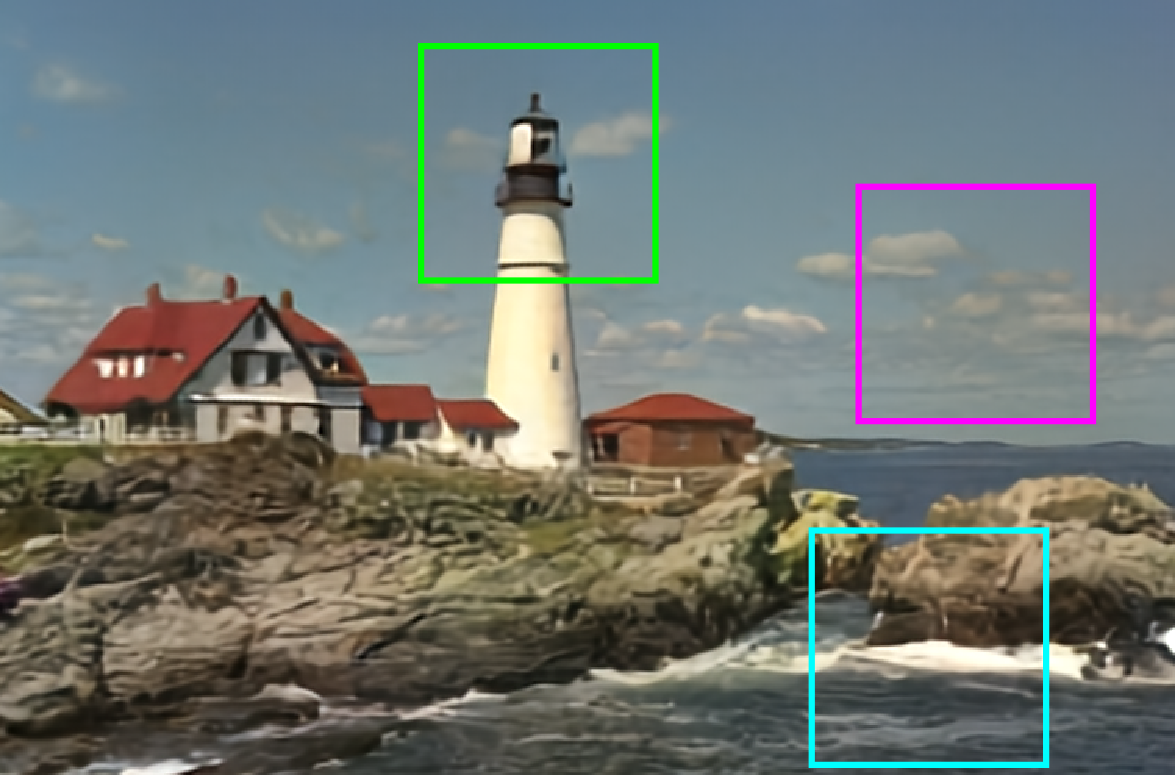}\vspace{2pt}
				\includegraphics[width=1\linewidth]{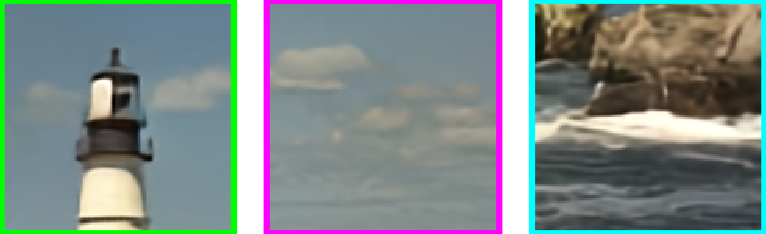}\vspace{-1pt}
				\centering {\small 0.933 / 0.124}
		\end{minipage}}
		\subfigure{
			\begin{minipage}[b]{0.185\linewidth}
				\centering {\small \bf Ours}\vspace{1pt}
				\includegraphics[width=1\linewidth]{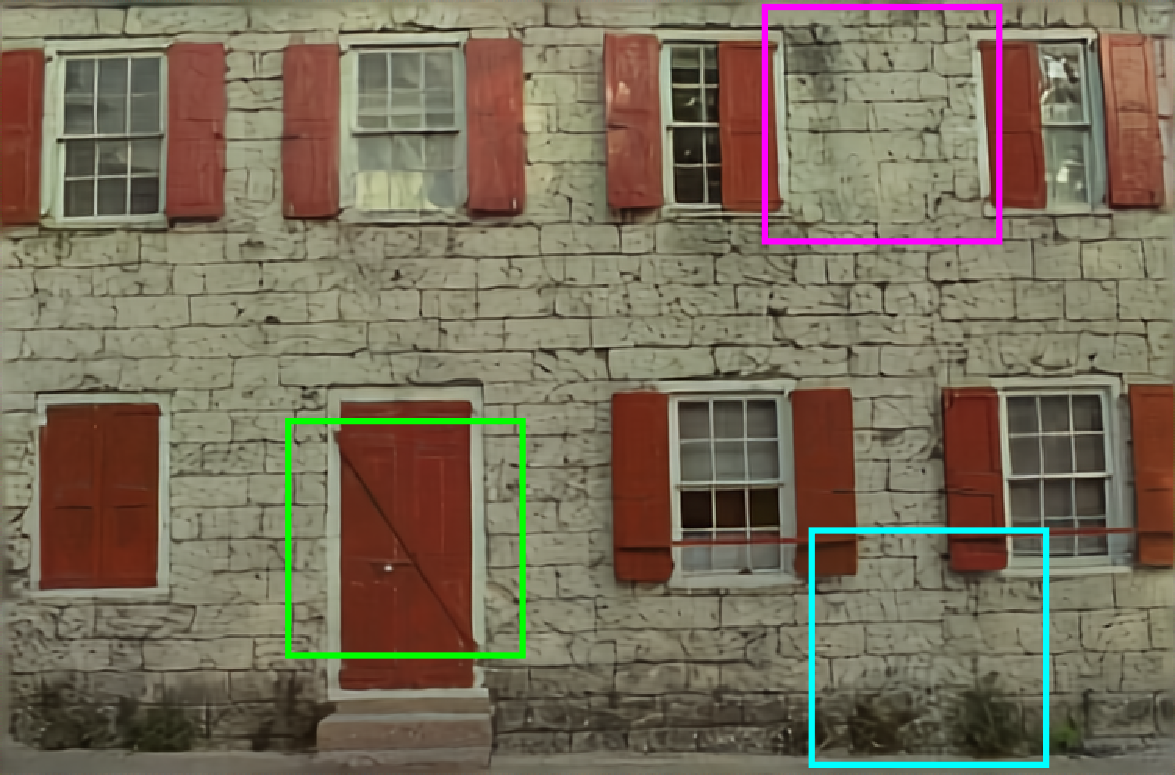}\vspace{2pt}
				\includegraphics[width=1\linewidth]{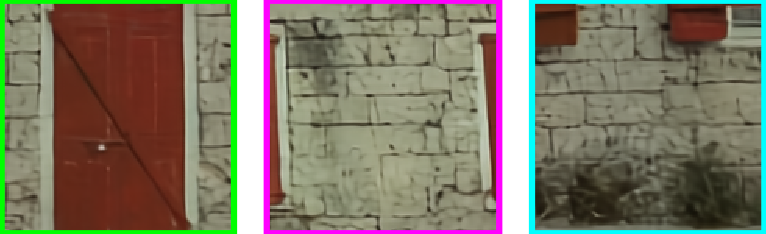}\vspace{-1pt}
				\centering {\small 0.952 / 0.242}\vspace{4pt}
				\includegraphics[width=1\linewidth]{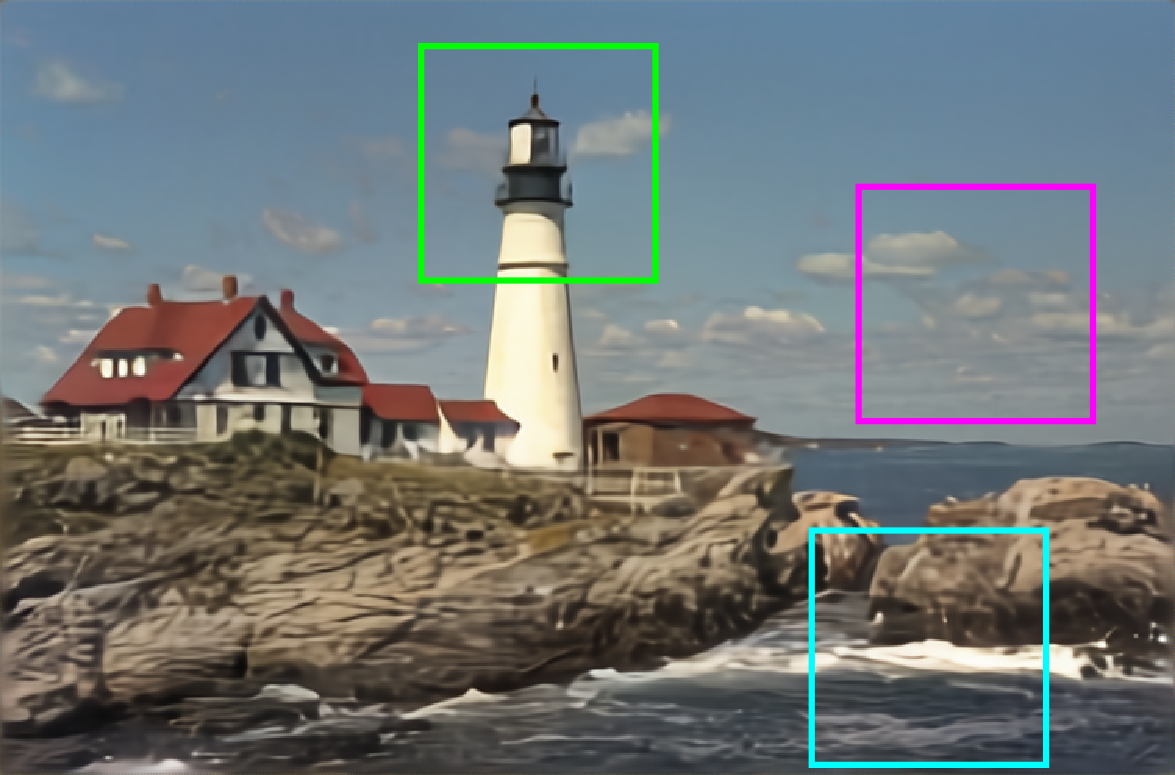}\vspace{2pt}
				\includegraphics[width=1\linewidth]{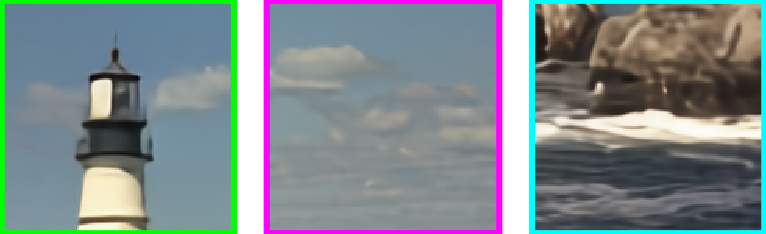}\vspace{-1pt}
				\centering {\small 0.943 / 0.125}
		\end{minipage}}
		\vspace{-10pt}
		\caption{Visual comparisons on example images (top: \textit{kodim1}, bottom: \textit{kodim21}) from the Kodak dataset. From left to right: the original images, WebP, BPG, Mentzer's, and ours. Our model achieves the best visual quality, demonstrating the superiority of our model in preserving both sharp edges and detailed textures. \textit{Best viewed on screen.}}
		\vspace{-16pt}
		\label{fig:examples}
	\end{figure*}

	\subsubsection{Investigation of Channel Importance Module}

	To demonstrate the effectiveness of the variable quantization mechanism and the channel importance module, we design several comparative experiments to evaluate the reconstruction performance. The baseline model generated a quantized feature map with channel number $C=32$. The quantization level vector $\mathbf{q}=[5]$ indicates that there are no splitting and merging operations. Thus, this model just contains one group. By contrast, with the same setting $\mathbf{q}=[3, 5, 7]^\top$ and $\mathbf{r}$, we use three different types of the channel importance module mentioned in Sec.~\ref{cim}, i.e., Sequeze and excitation block~(SE)-based, reconstruction error~(RE)-based, and predefined. We train these four variants for 400 epochs under the same training setting. We run all experiments three times and record the best MS-SSIM on Kodak. The details of the average results are listed in Tab.~\ref{table:variable}. We observe that the channel importance module and the splitting-merging module make the system more effective (smaller BPP) and powerful (better MS-SSIM). Additionally, the predefined channel importance module distinctly outperforms SE and RE-based modules, even SE and RE-based modules are learnable and data-dependent.  This may be consistent with the network pruning research~\cite{rethinkpruning}: training predefined target models from scratch can have better performance than pruning algorithms under some conditions. We also visualize the quantized feature map of the predefined model in Fig.~\ref{fig:visual_compare}. Comparing it with Fig.~\ref{fig:channel} (top right), we can see that the channels containing much more profile and context information of the original image are allocated more bits in the new system.

	\vspace{-1pt}
	\subsubsection{Investigation of the Combination in $\mathbf{q}$}

	As mentioned in Sec.~\ref{controller}, if the $G$ groups satisfy $\mathbf{r}^\top{\rm{log_2}}(\mathbf{q}) < {\rm{log_2}}(Q)$, the variable quantization controller will provide a lower theoretical entropy upper-bound. Here, we explore what combination may have better performance. The baseline model only has one quantization level, i.e., $\mathbf{q} = [5]$. We extend it to three types of combinations:  $\mathbf{q} = [4, 5, 6]^\top$, $\mathbf{q} = [3, 5, 7]^\top$, and $\mathbf{q} = [2, 5, 8]^\top$. The ratio vectors of the three types of models are the same and equal to $[25\%, 50\%, 25\%]^\top$. Quantitatively, $\log_2(2)+\log_2(8) < \log_2(3)+\log_2(7)<\log_2(4)+\log_2(6) < 2\log_2(5)$, and the experimental results are consistent with the theoretical analysis. Additionally, we find that the odd quantized level may have better performances. Because the odd quantized level more likely contains a quantized value close to 0. This may meet the similar results in research related to network quantization~\cite{zhu2016trained}. If the quantization levels in $\mathbf{q}$ are too different, e.g., $[2, 5, 8]^\top$, the performance will degrade.

	\vspace{-2pt}
	\subsection{Comparisons}
	\vspace{1pt}

	In this subsection, we compare the proposed method against three conventional compression techniques, JPEG2000, WebP, and BPG (4:4:4), as well as recent deep learning-based compression work by \cite{icrnn3}, \cite{icml2017}, \cite{weighted}, and \cite{conditional}.  We use the best performing configuration we can find of JPEG 2000, WebP, and BPG. Trading off between the distortion and the compression rate, $\mathbf{q}$ is set to $[3,5,7]^\top$ in the following experiments.

	\vspace{-1pt}
	\subsubsection{Quantitative Evaluation}
	\vspace{2pt}

	Following \cite{icml2017,conditional}, and because MS-SSIM is more consistent with human visual perception than PSNR, we use MS-SSIM as the performance metric. Fig.~\ref{fig:visual_compare} depicts the rate-distortion curves of these eight methods. Our method outperforms conventional compression techniques JPEG2000, WebP and BPG, as well as the deep learning-based approaches of \cite{icrnn2}, \cite{weighted}, \cite{conditional}, and \cite{icml2017}. This superiority of the proposed method holds for almost all tested BPPs, i.e., from 0.1 BPP to 0.6 BPP. It should be noted that both \cite{weighted} and \cite{conditional} are trained on the Large Scale Visual Recognition Challenge 2012 (ILSVRC2012)~\cite{imagenet2012}, which contains more than one million images. \cite{icml2017} trained their models on the Yahoo Flickr Creative Commons 100 Million dataset~\cite{thomee2016yfcc100m}, which includes approximately 100 million images. While our models are trained using only 4,000 images.

	\vspace{-1pt}
	\subsubsection{Visual Quality Evaluation}
	
	Owing to the lack of reconstruction results for many deep image compression algorithms and the space limitations of the paper, we present only two reconstruction results of images and compare them with WebP, BPG, and \cite{conditional}. In the first row of Fig.~\ref{fig:examples}, our method accurately reconstructs more clear and textural details of objects, e.g., door and the stripes on the wall. Other results have blocking artifacts more or less. For the second reconstruction results, our method can obtain better visual quality on images of objects such as clouds and water waves. Notably, our method is the only one that succeeds in reconstructing the spire of a lighthouse. Furthermore, the MS-SSIM measurements are also better than other methods in similar BBP ranges.

	\vspace{-6pt}
	\section{Conclusion}
	\vspace{1pt}
	
	In this paper, we propose, to the best of our knowledge, the first channel-level method for deep image compression. Moreover, based on the channel importance module and the splitting-merging module, the entire system can variably allocate different bitrates to different channels, which can further improve the compression rates and performances. Additionally, we formulate the quantizer into a GMM manner, which is a universal pattern for the existing finite range quantizers.  Ablation studies validate the effectiveness of the proposed modules. Extensive quantitative and qualitative experiments clearly demonstrate that our method achieves superior performance and generates better visual reconstructed results than the state-of-the-art methods without a hyperprior VAE.
	
	\vspace{-5pt}
	\section*{Acknowledgments} 
	This work was partially supported by JST CREST JPMJ-CR1686, Japan.
	
	\clearpage
	\small
	\bibliographystyle{named}
	\bibliography{ijcai20}
	
	\newpage
	\onecolumn
	\appendix
	\begin{center}
	{\LARGE{\textbf{Channel-Level Variable Quantization Network for Deep Image Compression}\\ \vspace{10pt} (Supplementary Material)}}
\end{center}

\vspace{20pt}

\section{Detail of Context Entropy Model}
\begin{figure}[h]
	\centering
	\begin{minipage}[c]{0.4\textwidth}
		\scalebox{1}[1] {\includegraphics[width=0.99\textwidth]{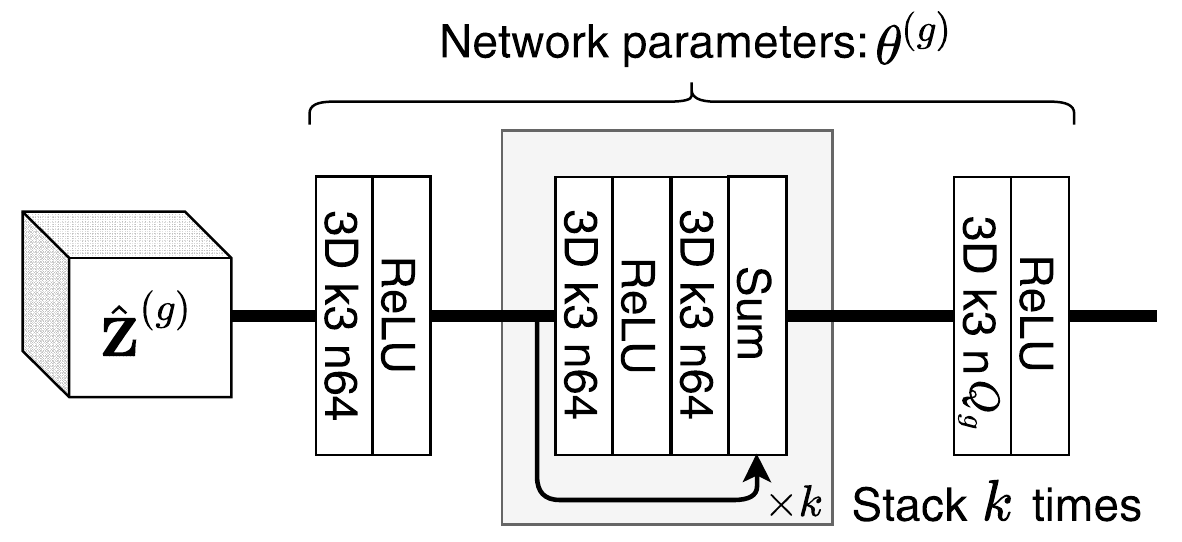}}
		\caption{Architecture of the context model copied from Mentzer's paper. ``3D k3 n64'' refers to a 3D masked convolution with filter size 3 and 64 output channels. The last layer outputs $Q_g$ values for each voxel in $\hat{\mathbf{Z}}^{(g)}$.}
		\label{fig:3dconv}
	\end{minipage}
	\hspace{15pt}
	\begin{minipage}[c]{0.52\textwidth}
		\centering
		\renewcommand\arraystretch{1.15}
		%
		
		\begin{tabular}{p{1.4cm}<{\centering}p{1.4cm}<{\centering}p{2.5cm}<{\centering}p{2.5cm}<{\centering}}
			\toprule 
			$\mathbf{q}^{\top}$ & CI Type   & MS-SSIM / BPP ($k=1$) & MS-SSIM / BPP ($k=3$)\\
			\hline \hline
			
			$[5]$ &  None  & 0.9651 / 0.2664 & 0.9651 / 0.2595\\
			\rowcolor{mygray}
			$[3, 5, 7]$ & SE-based & 0.9646 / 0.2608 & 0.9647 / 0.2587 \\
			
			$[3, 5, 7]$& RE-based & 0.9652 / 0.2586   & \textbf{0.9653} / 0.2571 \\
			\rowcolor{mygray}
			$[3, 5, 7]$ & Predefine & \textbf{0.9653} / \textbf{0.2576}  & 0.9652 / \textbf{0.2524}\\
			\bottomrule 
		\end{tabular}
		\captionof{table}[foo]{Performance of the 3D context entropy model with different $k$. Evaluation by MS-SSIM and BPP.}
		\label{tab:dk}
	\end{minipage}
\end{figure}

\vspace{10pt}
The loss function Eqn.~(\ref{eqn:loss}) contains three terms. The second term is the entropy loss of the quantized map $\hat{\mathbf{Z}}$, which is defined as $L_{{\rm{ent}},g}=-\sum \log p(\hat{\mathbf{Z}}^{(g)}| \mathbf{\theta}_g)$ and optimized by the context entropy model's parameters $\mathbf{\theta}_g$. $\mathbf{\theta}_g$ represents the $g$-th 3D CNN based context model (please refer to \cite{conditional} for detail) and its illustration is showing in Fig.~\ref{fig:3dconv}. We also evalute the performance of the 3D context entropy model with different $k$. From Tab.~\ref{tab:dk}, we can find that stacking more residual layers (larger $k$) can reduce BPP.

\section{Comparison on Other Datasets}

\vspace{10pt}

\begin{figure*}[h]
	\centering
	\begin{multicols}{2}
		\centering \includegraphics[width=0.45\textwidth]{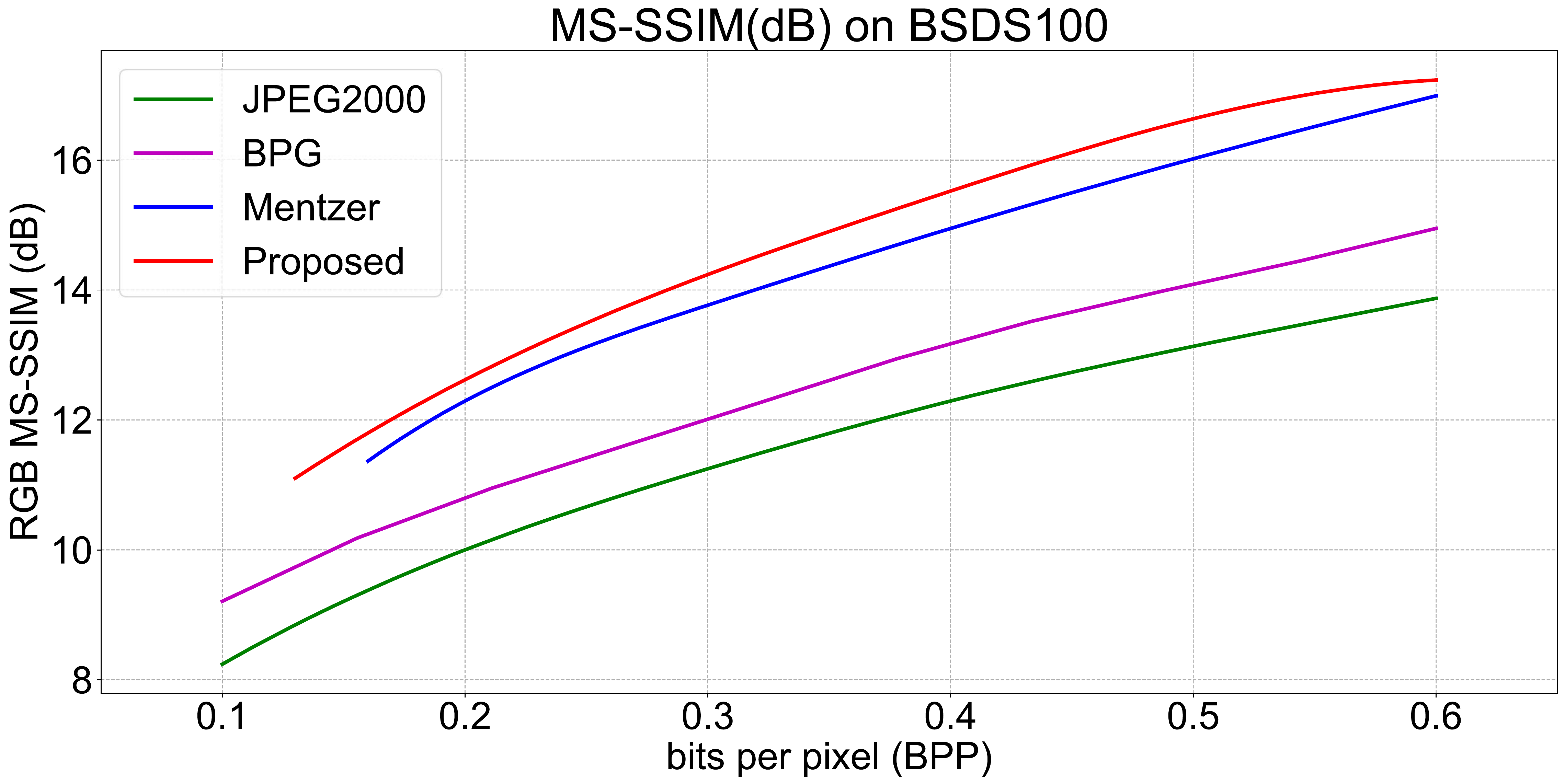} \hspace{15pt}\\
		\includegraphics[width=0.45\textwidth]{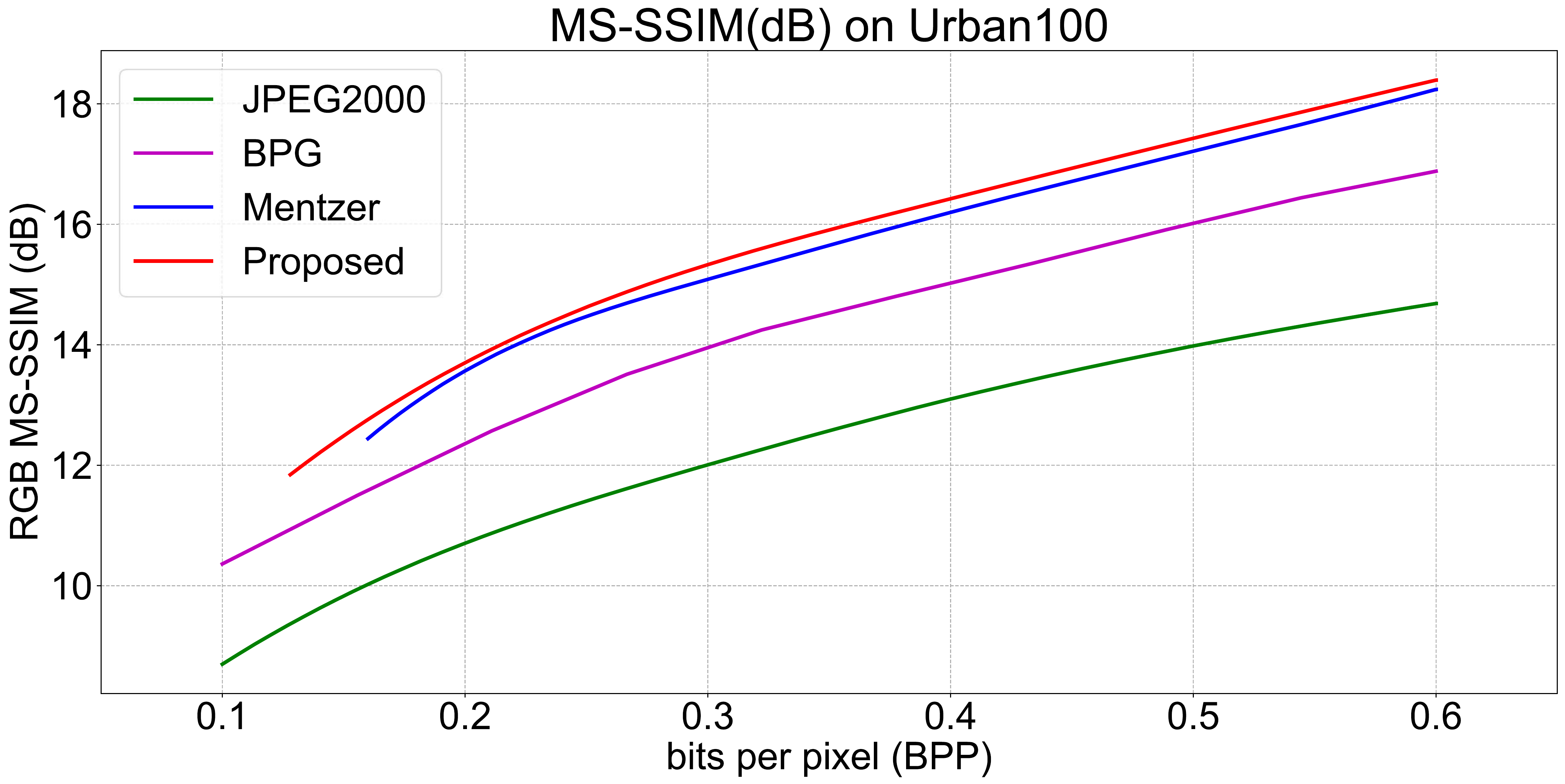}\\ 
	\end{multicols}
	\caption{Performance of our method on the BSDS100 dataset (left) and the Urban100 dataset (right), where we outperform Mentzer's, BPG and JPEG2000 for all tested BPPs, i.e., from 0.1 BPP to 0.6 BPP in MS-SSIM. \textit{Best viewed on-screen.}}
	\label{fig:visual_more}
\end{figure*}

Furthermore, to assess performance on high-quality full-resolution images, we test our proposed methods on the datasets BSDS100 and Urban100, commonly used in the super-resolution task. The experiment results are shown in Fig.~\ref{fig:visual_more}. Our method outperforms BPG and JPEG2000, as well as the neural network-based approach of \cite{conditional} for all tested BPPs, i.e., from 0.1 BPP to 0.6 BPP.

\clearpage

\section{PixelShuffle and Inverse PixelShuffle}

\vspace{10pt}

\begin{figure}[h]
	\begin{center}
		\includegraphics[width=0.7\linewidth]{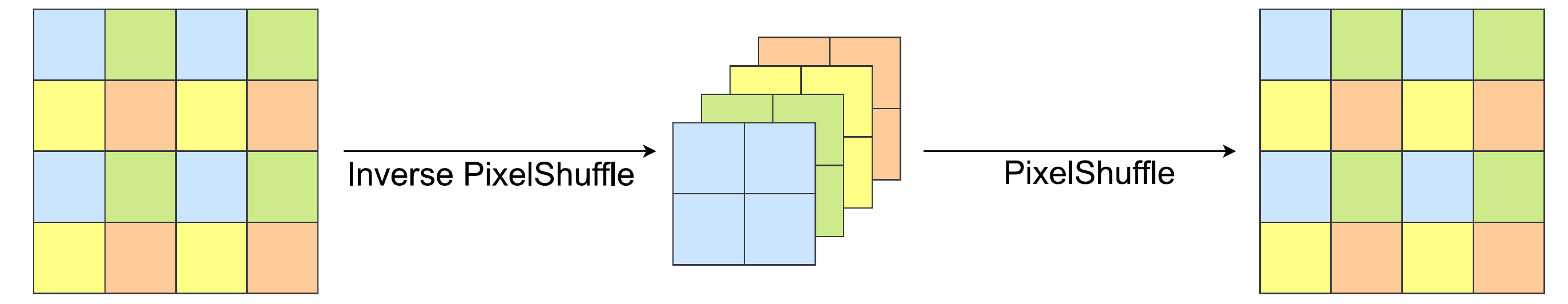}
	\end{center}
	\caption{Illustration of PixelShuffle and Inverse PixelShuffle.}
	\label{fig:ps}
\end{figure}
\vspace{-5pt}
\begin{equation}\label{eqn:ps}
\mathcal{PS}(\mathbf{X})_{c, h, w} = \mathbf{X}_{c+C \cdot {\rm{mod}}(h,d)+C \cdot {\rm{mod}}(w,d), \lfloor  h/d \rfloor, \lfloor  w/d \rfloor}.
\end{equation}
\vspace{3pt}
\begin{equation}\label{eqn:ips}
\mathcal{IPS}(\mathbf{X})_{c(di+j), h, w} = \mathbf{X}_{c, dh + i, dw + j}, \ 1\leq i,j \leq d.
\end{equation}
\vspace{3pt}

PixelShuffle~\cite{pixelshuffle}, i.e., Eqn.~(\ref{eqn:ps}), and Inverse PixelShuffle, i.e., Eqn.~(\ref{eqn:ips}), are very simple operations for down-sampling and up-sampling. We think these two operations are very suitable for deep image compression because they can vary the spatial resolution of the feature map and reconstruct the original input without information loss. (This may share some similar insight with residual learning in ResNet.) Moreover, comparing with down-sampling and up-sampling convolutions, there are more efficient both for computation and memory. In our experiments, we also found that inverse PixelShuffle can improve the stability of training while sometimes down-sampling convolution (stride $>$ 1) fails to converge. Last, PixelShuffle is widely used in the super-resolution task~\cite{rcan,srcliquenet}, which also indicates this simple operation is effective for resolution transform.

\vspace{10pt}

\section{Exploration of different $G$ and $\mathbf{r}$}

\vspace{10pt}

Here we conduct a comparison experiment. We design three variants by varying $G$, $\mathbf{r}$ and $\mathbf{q}$. All variants are trained for 200 epochs under the same setting and evaluated on the Kodak dataset. Tab.~\ref{table:vary_g} shows the detailed results. From it, the trend of BPPs almost follows the values of $\mathbf{r}^\top\log_2\mathbf{q}$, which is consistent with our analysis in Sec.~\ref{controller}. As $G$ becomes larger, BPP can reduce further with negligible loss of MS-SSIM.  However, as $G$ becomes larger, the number of channels in each segment will decrease, which may influence the performance of the context
entropy model due to the relevant or dependent information
among the channels is reduced. Additionally, we found that increasing $G$ will make the training of the whole deep compression system unstable.

\begin{table*}[h]
	\renewcommand\arraystretch{1.2}
	\centering
	\setlength{\tabcolsep}{0.3mm}{
		\begin{tabular}{p{1.3cm}<{\centering}p{3cm}<{\centering}p{2.3cm}<{\centering}p{2.3cm}<{\centering}p{2.3cm}<{\centering}p{2.3cm}<{\centering}p{2.3cm}<{\centering}p{2.3cm}<{\centering}}
			\toprule 
			$G$      & $\mathbf{r}^\top$    & $\mathbf{q}^\top$    & $\mathbf{r}^\top\log_2\mathbf{q}$  &PSNR   & MS-SSI$\rm{M}_{dB}$ & BPP \\
			\hline \hline
			\rowcolor{mygray}
			2 & [1/2, 1/2] & [4, 6] & 2.293  & 27.581 & 14.282 & 0.2570\\
			3 & [1/3, 1/3, 1/3] & [3, 5, 7] & 2.238  & 27.422 & 14.272 &0.2518\\
			\rowcolor{mygray}
			4 & [1/4, 1/4, 1/4, 1/4] & [2, 4, 6, 8] &  2.146 &  27.383 & 14.152 &0.2431\\
			\bottomrule 
	\end{tabular}}
	
	\caption{Investigation of $G$, $\mathbf{r}$ and  $\mathbf{q}$. The channels number of quantized feature map $C$ equals to $32$ for all experiments.}
	\label{table:vary_g}
	
\end{table*}

\clearpage

\section{More Visualization Results}
In this section, we provide a few more visualization reconstructed results at a wider BPP range. All figures include the ground truth images (left) and the reconstructed images for BPG  (middle) and our proposed method (right).

\begin{figure*}[h!]
	\centering
	\begin{multicols}{3}
		\centering \includegraphics[width=0.32\textwidth]{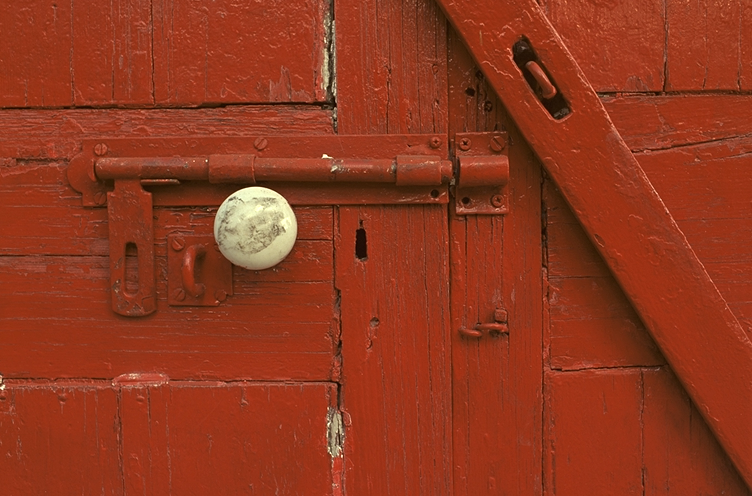} 
		\centering { MM-SSIM / MM-SSIM (dB) / BPP}\hspace{15pt} \\
		\includegraphics[width=0.32\textwidth]{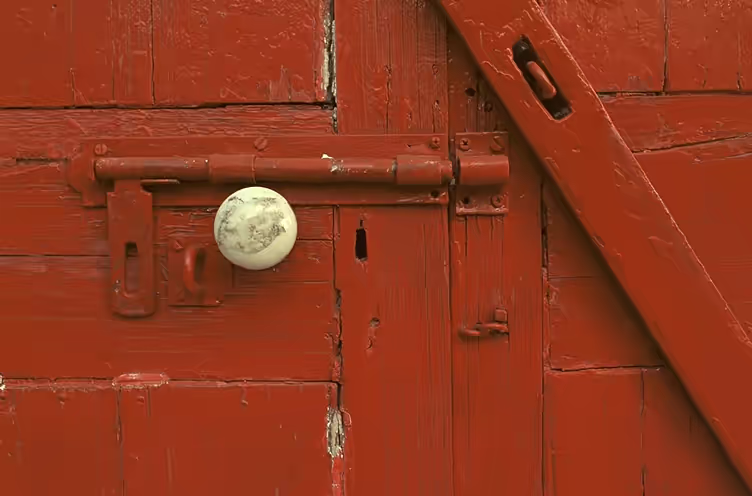}\\ 
		\centering {BPG: 0.951 / 13.09 / 0.280} \hspace{15pt} \\
		\includegraphics[width=0.32\textwidth]{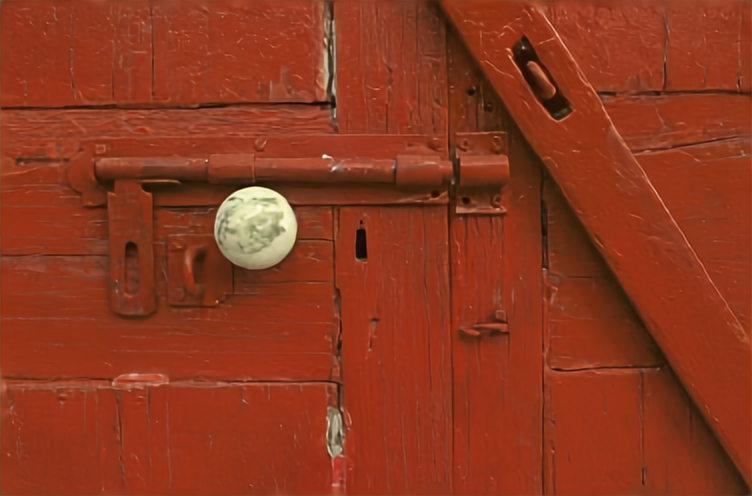}\\
		\centering {\textbf{Ours: 0.964 / 14.44 / 0.262}}
	\end{multicols}
	\begin{multicols}{3}
		\centering \includegraphics[width=0.32\textwidth]{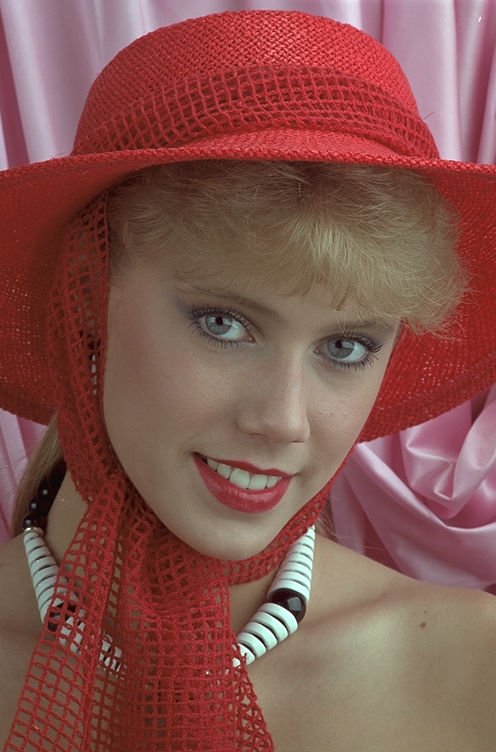} \centering { MM-SSIM / MM-SSIM (dB) / BPP}\hspace{15pt} \\
		\includegraphics[width=0.32\textwidth]{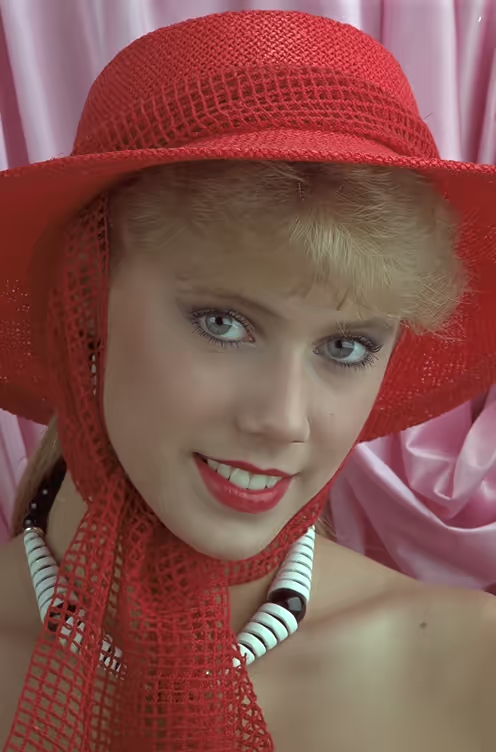}\\ 
		\centering {BPG: 0.970 / 15.23 / 0.399}  \hspace{15pt} \\
		\includegraphics[width=0.32\textwidth]{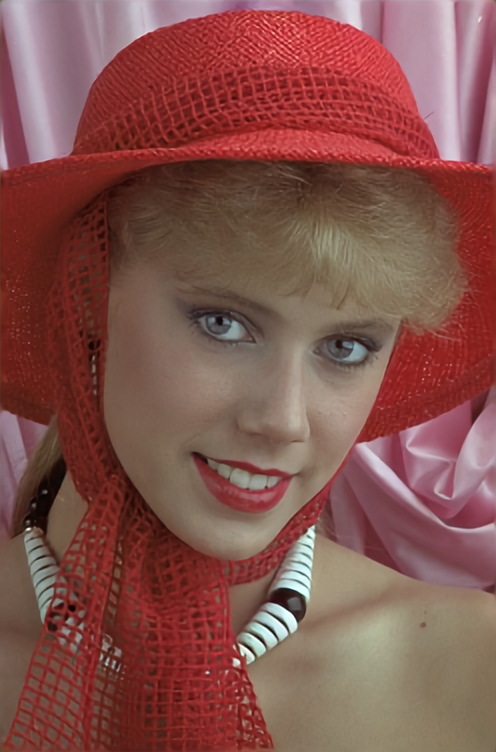}\\
		\centering { \textbf{Ours: 0.975 / 16.02 / 0.387}}
		
	\end{multicols}
	\begin{multicols}{3}
		\centering \includegraphics[width=0.33\textwidth]{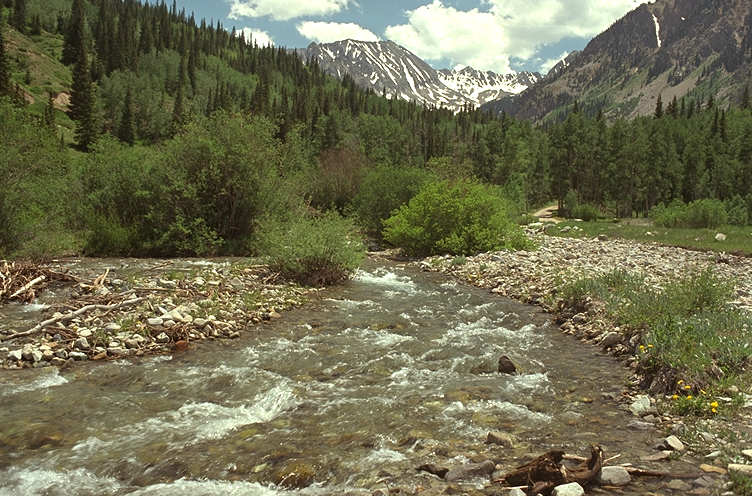} \centering { MM-SSIM / MM-SSIM (dB) / BPP}\hspace{15pt} \\
		\includegraphics[width=0.33\textwidth]{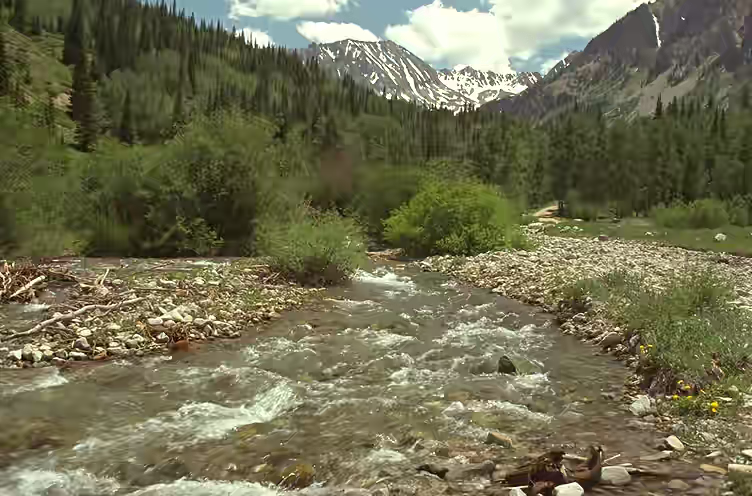}\\ 
		\centering {BPG: 0.931 / 11.61 / 0.583} \hspace{15pt}\\
		\includegraphics[width=0.33\textwidth]{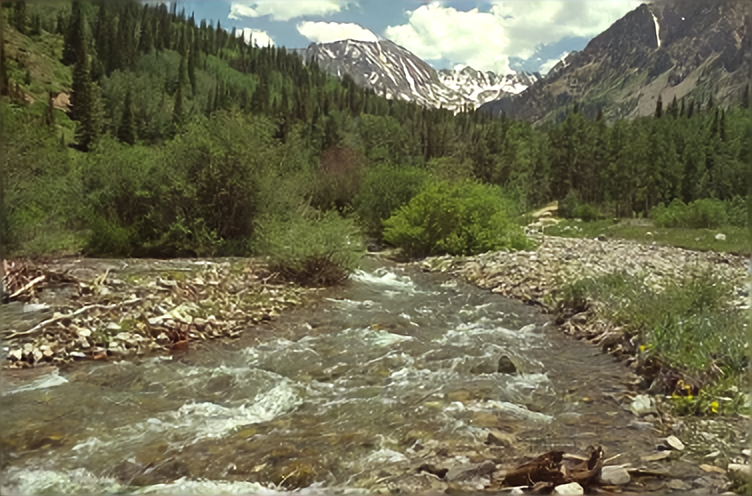} \\
		\centering {\textbf{Ours: 0.965 / 14.55 / 0.534}}
		
	\end{multicols}
	\caption{Some sample test results from the Kodak dataset (\textit{kodim2}, \textit{kodim4} and \textit{kodim13}). At similar bit rates, our combined method provides the highest visual quality. BPG shows more “classical” compression artifacts. \textit{Best viewed on-screen.}}
	\label{fig:vis}
\end{figure*}

\end{document}